\documentclass[lettersize,journal]{IEEEtran}
\usepackage{amsmath,amsfonts}
\usepackage{algorithmic}
\usepackage{array}
\usepackage[caption=false,font=normalsize,labelfont=sf,textfont=sf]{subfig}
\usepackage{textcomp}
\usepackage{stfloats}
\usepackage{url}
\usepackage{verbatim}
\usepackage{graphicx}
\usepackage{amsmath}
\usepackage{makecell}
\usepackage{cite}
\usepackage{balance}
\usepackage{tabu}                      
\usepackage{booktabs}                  
\usepackage{lipsum}                    
\usepackage{mwe}                       
\usepackage{multirow}
\usepackage[normalem]{ulem}
\useunder{\uline}{\ul}{}
\usepackage{caption}
\usepackage{float}
\usepackage{orcidlink}
\usepackage{algorithm}

\usepackage[dvipsnames,svgnames]{xcolor}
\usepackage[most]{tcolorbox}
\usepackage{graphicx}
\usepackage{enumitem}
\usepackage{setspace}
\setlist{noitemsep,parsep=0pt,partopsep=0pt} 
\newcommand{\eg}{e.g.}
\newcommand{\ie}{i.e.}
\newcommand{\etal}{et al.}
\newcommand{\unit}{stage}
\newcommand{\units}{stages}
\newcommand{\behaviorunit}{behavior stage}
\newcommand{\behaviorunits}{behavior stages}
\newcommand{\activebehavior}{proactive behavior}
\newcommand{\activebehaviors}{proactive behaviors}
\newcommand{\passivebehavior}{reactive behavior}
\newcommand{\passivebehaviors}{reactive behaviors}
\newcommand{\systemname}{\textit{\textit{EMINDS}}\xspace}
\newcommand{\behaviourclusterview}{Behavior Cluster View\xspace} 
\newcommand{\progressionstageview}{Behavior Stage View\xspace} 
\newcommand{\progressionview}{Behavior Progression View\xspace} 
\newcommand{\patternview}{Pattern View\xspace} 
\newcommand{\newSankey} {pattern-centric Sankey diagram\xspace}

\newcommand{\depressiondegree} {\textit{depression degree}}

\newcommand{\Ea} {\textit{$E_{a}$}\xspace}
\newcommand{\Eb} {\textit{$E_{b}$}\xspace}
\newcommand{\Ec} {\textit{$E_{c}$}\xspace}
\newcommand{\Pa} {\textit{$P_{a}$}\xspace}
\newcommand{\Pb} {\textit{$P_{b}$}\xspace}
\newcommand{\Pc} {\textit{$P_{c}$}\xspace}
\newcommand{\Pd} {\textit{$P_{d}$}\xspace}

\definecolor{verylightgray}{gray}{0.9}

\newcommand{\typeiconp}{\hspace{-0.6mm}\tikz\draw[fill=verylightgray, draw=black] (0,0) circle (.9ex);\hspace{0.6mm}}

\newcommand{\typeiconr}{\hspace{-0.5mm}{\tikz\draw[fill=verylightgray, draw=black] (0,0) -- (0.25cm,0) -- (0.125cm,0.23cm) -- cycle;}\hspace{0.5mm}}

\newcommand{\typeiconm}{\hspace{-0.5mm}{\tikz\draw[fill=verylightgray, draw=black] (0,0) rectangle (0.25cm,0.25cm);}\hspace{0.5mm}}

\usepackage{mathptmx}                  
\usepackage{CJK}

\definecolor{border}{RGB}{100, 100, 100}
\newtcbox{\mybox}[1][border]
  {on line, arc = 1pt, outer arc = 1pt, colframe = border,
    colback = #1!4!white, boxsep = 0pt, left = 1pt, right = 1pt, top = 1pt, bottom = 1pt, boxrule = 1pt}

\AtBeginDocument{

}
\definecolor{highlight}{HTML}{1155cc}
\definecolor{highlight}{HTML}{000000}

\newcommand{\rui}[1]{\textcolor{highlight}{#1}}

\definecolor{review}{HTML}{f72585}
\usepackage{marginnote}
\usepackage{adjustbox}
\global\marginparsep=9pt
\newcommand{\sidecomment}[1]{%
  \ifdefined\revise
  \marginnote{%
    \textcolor{review}{%
      \adjustbox{minipage=1.1\marginparwidth,fbox}{%
          \scriptsize#1%
      }
    }
  }
  \fi
}

\def\BibTeX{{\rm B\kern-.05em{\sc i\kern-.025em b}\kern-.08em
    T\kern-.1667em\lower.7ex\hbox{E}\kern-.125emX}}
\usepackage{balance}
\begin{document}
\title{\textit{EMINDS}: Understanding User Behavior Progression for Mental Health Exploration on Social Media}
\author{
Rui Sheng\orcidlink{0000-0001-9321-6756},
Yifang Wang$^{*}$\orcidlink{0000-0001-6267-9440},
Xingbo Wang\orcidlink{0000-0001-5693-1128},
Shun Dai\orcidlink{0009-0001-1542-379X},
Qingyu Guo\orcidlink{0000-0002-8509-3933},\\
Tai-Quan Peng\orcidlink{0000-0002-2588-7491},
Huamin Qu\orcidlink{0000-0002-3344-9694},
and Dongyu Liu$^{*}$\orcidlink{0000-0002-8915-2785}

\vspace{-1em}

\thanks{This paper was produced by the IEEE Publication Technology Group. They are in Piscataway, NJ.}
\thanks{Manuscript received April 19, 2021; revised August 16, 2021.}
\thanks{Rui Sheng, Qingyu Guo, Huamin Qu are with the Hong Kong University of Science and Technology (rshengac@connect.ust.hk, qguoag@connect.ust.hk, huamin@cse.ust.hk); Yifang Wang is with Florida State University (yifang.wang@fsu.edu); Xingbo Wang is with Cornell University  (xingbo.wang@med.cornell.edu); Shun Dai is with the University of Tokyo (daishun0102@gmail.com); Tai-Quan Peng is with Michigan State University (pengtaiq@msu.edu), Dongyu Liu is with the University of California, Davis (dyuliu@ucdavis.edu)}
\thanks{$^{*}$Co-corresponding authors}
\thanks{The paper has been accepted for publication in IEEE Transactions on Visualization and Computer Graphics, with a DOI of 10.1109/TVCG.2025.3630646.}
\thanks{© 20XX IEEE.  Personal use of this material is permitted.  Permission from IEEE must be obtained for all other uses, in any current or future media, including reprinting/republishing this material for advertising or promotional purposes, creating new collective works, for resale or redistribution to servers or lists, or reuse of any copyrighted component of this work in other works.}
}

\markboth{Journal of \LaTeX\ Class Files,~Vol.~14, No.~8, August~2021}%
{How to Use the IEEEtran \LaTeX \ Templates}
\IEEEpubid{0000--0000/00\$00.00~\copyright~2021 IEEE}

\maketitle

\begin{abstract}
Mental health is an urgent societal issue, and social scientists are increasingly turning to online mental health communities (OMHCs) to analyze user behavior data for early intervention. 
However, existing sequence mining techniques fall short of the urgent need to explore the behavior progression of different groups (e.g., recovery or deterioration groups) and track the potential long-term impact of behaviors on mental health status.
To address this issue, we introduce \systemname, a visual analytics system built on a novel automatic mining pipeline that extracts distinct behavior stages and assesses the potential impact of frequent stage patterns on mental health status over time.
The system includes a set of interactive visualizations that summarize the meaning of each behavior stage and the evolution of different stage patterns. 
We feature a \newSankey to reveal contextual information about the impact of stage patterns on mental health, helping experts understand the specific changes in sequences before and after a stage pattern.
We evaluated the effectiveness and usability of \systemname through two case studies and expert interviews, which examined the potential stage patterns impacting long-term mental health by analyzing user behaviors on Reddit.
\end{abstract}


\section{Introduction}
\IEEEPARstart{M}{ental} health issues, such as depression and eating disorders, significantly impact well-being. 
A recent survey~\cite{mhinus} shows that over 60 million U.S. adults are affected by mental illness, with 12.8 million having serious suicidal thoughts.
Therefore, early detection and intervention are crucial for social scientists in the healthcare domain.
\rui{
Online mental health communities (OMHCs), such as Reddit\footnote{https://www.reddit.com/r/depression/}, offer open access to rich behavioral data~\cite{naslund2020social}.}
\sidecomment{R1C1}
Users post messages to seek help and interact with others (\autoref{fig: concept}-A), providing valuable resources for social scientists who study mental health and beyond to explore topics like communication and human behaviors.
Based on these online behaviors, social scientists can derive different measures as behavior events (\eg, posting frequency and the number of responses received (\autoref{fig: concept}-B)) within each day or a specific time unit.
Then, event sequences (\autoref{fig: concept}-C), consisting of events over a long time period (such as over a year), enable researchers to study user behavior dynamics.
Specifically, social scientists often interpret multiple events collectively in event sequences as a \textbf{behavior stage} (\autoref{fig: concept}-D) to gain temporal insights~\cite{liu2017assessing}.
These behavior stages serve as meaningful units (\eg, consistently high posting frequency with receiving few responses) to characterize the progression of user groups (\autoref{fig: concept}-E), enabling the detection of intricate patterns that influence mental health in the long term~\cite{liu2017assessing, kushner2020bursts, peng2020mobile, ernala2021social}.
These findings can aid policymakers in mental health resource allocation and guide preventive strategies.

\IEEEpubidadjcol
However, existing works still have limitations in analyzing the progression of behavior stages for a user group. 
First, they fail to consider the interactions between users' diverse behaviors. For example, users may share personal updates and seek support, while also receiving posts from others and gaining social support. These interactions lead to different behavior stages.
Second, prior studies often ignore the influence of mental health status on behavior. However, mental health might significantly impact online behaviors~\cite{akhther2022seeking}.
For instance, individuals with poor mental health, during certain periods, exhibit a greater inclination to post online~\cite{de2013predicting}.
Ignoring this factor can lead to incomplete analyses and hinder effective intervention strategies.

While automated algorithms, such as segmentation and clustering approaches~\cite{hallac2017toeplitz, hallac2016ggs}, can distill behavior stages from multivariate event sequences, they pose significant challenges for experts attempting to interpret the results.
The first challenge is that each behavior stage is represented by a cluster consisting of numerous similar event sequences with varied lengths, which makes it difficult to summarize the meaning. 
The second challenge stems from the high number of involved variables and their complex interactions.
Visual analytics has been shown to be useful in helping domain experts analyze event progression~\cite{wu2021tacticflow, wang2021threadstates, kwon2020dpvis}.
However, previous works primarily focus on summarizing the varied-length sequences of events that involve only one variable \cite{visual2018guo, eventthread2018guo}.
Moreover, social scientists need to detect frequent stage patterns and summarize their long-term impact on mental health.
Unfortunately, prior studies ~\cite{wu2021tacticflow, wang2021threadstates, kwon2020dpvis} cannot adequately enable them to comprehend how a stage pattern can result in changes in behaviors before and after.

To overcome these limitations, we introduce \systemname, a visual analytics system designed to empower experts in exploring vast quantities of multivariate behavior sequences.
Specifically, we designed an automatic mining pipeline for user behavior analysis, which can help researchers identify behavior stages, extract frequent stage patterns, and analyze their impact on mental health.
To enhance experts' understanding of the results generated by the automated pipeline, we proposed a set of interactive visualizations to help them summarize and explore the implications of each behavior stage. 
We also introduced a pattern-centric Sankey diagram, which empowers experts to understand intricate information about the behavior change triggered by a specific stage pattern.
In summary, our contributions are as follows:
\begin{itemize}[leftmargin=1em]
    \item We formulate the problem for analyzing user behaviors and their impact on mental health within OMHCs. By adapting key concepts from social science and proposing new definitions, we provide a structured approach that enhances the understanding of complex user behavior progression.

    \item We design and implement a novel visual analytics system\footnote{https://github.com/dylansheng/EMINDS} (\systemname) that integrates an automatic mining pipeline with interactive visualization techniques. 
    The system empowers social scientists to extract, explore, and analyze behavior stages from large-scale multivariate event sequences. 
    We also provided a demo\footnote{https://www.tvcg-eminds.com/} with videos\footnote{https://github.com/dylansheng/EMINDS-video} to showcase our system.
    
    \item We prove the effectiveness of \systemname through expert interviews and in-depth case studies, uncovering valuable insights into the progression of user behaviors in OMHCs.
\end{itemize}
\section{Related Works}
\subsection{User Behavior Analysis in OMHCs}
OMHCs play a crucial role for people with mental health issues to find support~\cite{sharma2020engagement,gatos2021hci}.
Typically, users can start a post about their mental health issues, and community members will respond with support~\cite{peng2021effects}. 
A vast and varied amount of behavior data is generated on OMHCs, providing valuable insights for social scientists to study online intervention~\cite{naslund2020social,chancellor2020methods, Saha_Sharma_2020,peng2021effects}, especially by considering the progression of behavior stages.
For example, the expected types of social support could vary as users gain experience on OMHCs~\cite{peng2021effects}. 
While it has been suggested that users' mental health status, generated behaviors, and perceived support are all intertwined~\cite{akhther2022seeking}, previous studies have often overlooked their complex interactions. 
As a result, the overall progression is often poorly understood.
Moreover, social scientists need to explain and validate these findings to develop online intervention technologies. 
Therefore, we provide a visual analytics system to help experts analyze complex user behavior progression in mental health research.

\begin{figure*}[ht]
\centering
\includegraphics[width=1\linewidth]{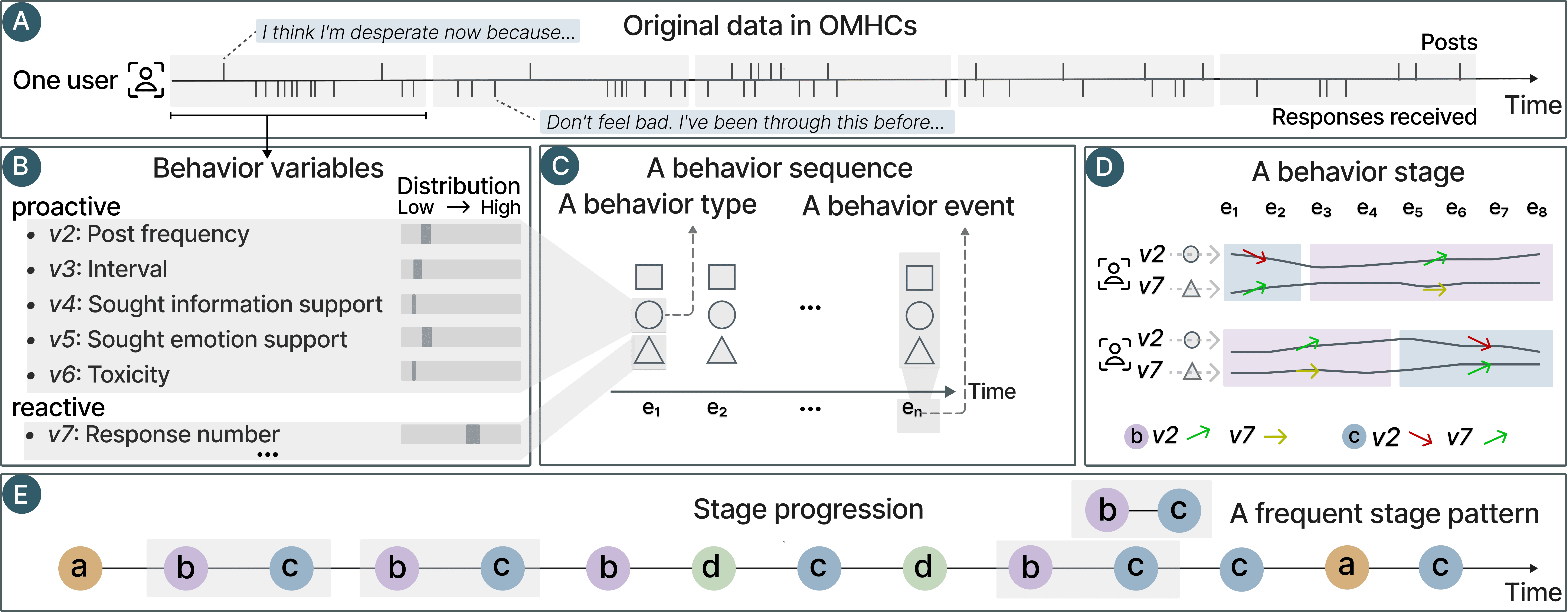}
\caption{The concepts we have adapted from social science or proposed to improve understanding of the problem. (A) The original user posting data in OHMCs is collected. (B) Behavior variables are derived from user posts. (C) Behavior types are extracted and combined to form behavior events, which will be used to form behavior sequences. (D) Behavior stages are extracted based on changes in behavior sequence over time. (E) Original behavior sequences are represented by stage progression to identify frequent stage patterns.}
\label{fig: concept}
\vspace{-1em}
\end{figure*}

\subsection{Progression Analysis of Event Sequences}\label{progression_rw} 
Event progression analysis is a long-standing task in various domains, aiming to uncover how a sequence of events (i.e., a sequence pattern) progresses over time.
There have been numerous automatic approaches, including statistical methods~\cite{hallac2017TICC, hallac2016ggs, Yang2014Finding} or machine learning methods~\cite{de2021change, deldari2021time}.
However, researchers often face challenges in fully exploring data progression using these methods, particularly for group-level analyses due to data complexity.

To enhance comprehension and potential control over automated algorithms, extensive research has focused on event sequence visualization methods. 
Guo \etal~\cite{guo2022survey} summarized the approaches to leverage visual analytics in progression analysis from different dimensions, including design space and visual analysis tasks.
Some works~\cite{kwon2016visohc, li2021conviscope, jin2021visual} visualize each event sequence directly, but this method becomes challenging when there are many sequences to analyze. Combining the same events~\cite{guo2019uncertainty, bartolomeo2021sequence, Zhang2015Iterative, Magallanes2022Sequen-C, Cappers2018Exploring, agarwal2020set} at a similar time for visualization is an approach to exploring numerous sequences. However, it is hard to study excessively long sequences over a period of time.

To overcome the challenge mentioned above, several works~\cite{wu2021tacticflow, wu2022rasipam, wang2021tacminer, wu2020visual, Blascheck2016VA2} have been adopting automatic methods to extract frequent patterns existing within an event sequence aggregation.
For example, TacticFlow~\cite{wu2021tacticflow} combined Generative Adversarial Networks (GANs) and visual analytics to help experts extract and understand multiple multivariate patterns in sports tactic analysis.
RASIPAM \cite{wu2022rasipam} strengthened the control over the algorithm's results by allowing experts to add constraints on the extracted patterns.
However, those works utilize frequent pattern mining to extract an exact ordered sequence of patterns to represent a stage.
It differs from ours, where a stage is composed of similar varied-length multivariate sub-sequences instead of frequent patterns.

Other studies leveraged similarity measures or transfer probabilities to classify events into different latent stages, which represent clusters of similar events. 
For example, ThreadStates \cite{wang2021threadstates} and DPVis \cite{kwon2020dpvis} mapped all the events into several stages through K-means and hidden Markov models, respectively.
However, they did not computationally summarize events over time within latent stages, which is important for studying mental health to gain temporal insights~\cite{huang2017highresolution}.
Other papers~\cite{visual2018guo, eventthread2018guo} proposed stage-based progression summarization approaches based on summarizing varied-length sub-sequences.
However, they focus on analyzing univariate event sequences, which only need to consider a single variable of an event.
To address these challenges, we first introduce an automatic mining pipeline that extracts behavior stages by clustering variable-length multivariate sequences. We also offer a practical approach to help experts understand the semantic meanings of each stage through our designed visualizations.

\subsection{Visual Analytics for Event Sequences in Social Media} 
Understanding user behaviors on social media through visual analytics has received rising attention~\cite{wu2016survey,guo2022survey}.
One area of research focuses on exploring ego-centric behaviors~\cite{cao2016episogram, cao2016targetvue, chen2020viseq}.
Those studies enable experts to analyze individual behaviors.
For example, TargetVue~\cite{cao2016targetvue} proposed an approach to support experts in monitoring unusual communication behaviors.
Other studies prioritized analyzing collective behaviors through visual analytics~\cite{li2020weseer, chen2021cobridges, chen2020rmap, law2019maqui}.
\sidecomment{R3C3}
\rui{When examining the temporal progression of behaviors, aggregating multiple behavior events has been proposed to reduce the difficulty of interpreting overly long sequences. Combining behavior events in this way can also help users better interpret their semantic meanings, especially in the case of multi-variate behavior events such as those in our scenario. 
Similar to the approaches of event sequence progression analysis in \autoref{progression_rw}, these methods still fail to address our problem.
Specifically, some methods~\cite{cao2016targetvue, Cho2017CrystalBall} segment the data based on a fixed time period, but this rigid slicing may obscure the dynamic evolution of behavior over time.
Frequent pattern mining is another approach~\cite{chen2020viseq} for aggregation, but it is less suitable in our context since social scientists are not strict about the exact order or length of patterns. An alternative is to summarize event sequences into stages by segmenting them according to their distinct trends and clustering similar segments, which has been used in event progression visualization and can overcome the above limitations~\cite{kwon2020dpvis, wang2021threadstates}. However, little research has explored how to effectively summarize and visualize stages from event sequences when each behavior event consists of multiple variants. Thus, the key challenge we address is how to develop methods that can generate and visualize meaningful stage-based summaries of multi-variate behavior events.}


\section{Background}\label{background}
We aim to develop a visual analytics system to help social scientists explore user behaviors associated with their mental health status for early intervention.
\sidecomment{R2C1, R3C1}
\rui{Over the past year, we have closely worked with two social scientists (\Ea, \Eb) and a doctoral student (\Ec). 
\Ea is a professor who has over two decades of mental health and social media research experience, currently conducting research in the United States.
\Eb is a professor who has six years of experience in the same field in Singapore. 
\Ec has worked on social support in OMHCs for four years in China.
\Ea is our co-author.}
We consulted with the experts to establish and iterate research goals.
We kept refining our research objectives through regular meetings with \Ea and \Ec twice a week and bi-weekly meetings with \Eb.
We will first introduce several key concepts in user behavior analysis. Following that, we will elaborate on the analysis goals and challenges of experts and outline the five design requirements.

\subsection{Key Concepts}\label{expert}
During our collaboration, we adapted concepts from social science (\textit{italic}) and proposed several new definitions (\textbf{\textit{bold italic}}) to formally define the problem (\autoref{fig: concept}). 
We define a \textit{\textbf{behavior event}} (\autoref{fig: concept}-C) as comprising the user’s mental health status (\hspace{1mm}\typeiconm), along with two \textit{\textbf{behavior types}}: a proactive behavior type (\hspace{1mm}\typeiconp) and a reactive behavior type (\hspace{1mm}\typeiconr) within a time unit.
This formulation provides comprehensive insights into the user's mental health status, their actions based on this status (\ie, proactive behaviors \cite{parker2010taking} such as posting), and the feedback received from others (\ie, reactive behaviors such as receiving responses \cite{rozo2016learning}).
Specifically, proactive behavior types are depicted by all the proactive \textit{behavior variables} with varying ranges or categories, while reactive behavior types are depicted by all the reactive behavior variables.
The following are the detailed descriptions.

\begin{itemize}[leftmargin=1em]
    \item \textit{Behavior variables} (\eg, \textit{v2} in \autoref{fig: concept}-B) are measured directly from social media data to characterize multiple aspects of user behaviors (\eg, post frequency and post interval in certain time periods)~\cite{De2021Predicting}. 
    These variables are classified into two categories: proactive behavior variables (initiated by users themselves, such as post frequency) and reactive behavior variables (passively experienced by users, such as the number of responses received). 
    
    \item \textit{\textbf{Behavior types}} (\eg, \typeiconp and \typeiconr
 in \autoref{fig: concept}-C) are derived from behavior variables and are also categorized into proactive and reactive behavior types. This categorization captures fundamental differences in user behaviors, helping experts interpret the role of each behavior variable from a higher-level perspective and alleviating the cognitive burden associated with interpreting an excessive number of individual behavior variables. Specifically, the former ( \typeiconp) is depicted by all the proactive behavior variables, while the latter ( \typeiconr) is depicted by all the reactive behavior variables. 
    Assigning each behavior variable to various ranges or categories allows the representation of different behavior types. For example, considering just two proactive behavior variables---post frequency (\ie, \textit{v2} in \autoref{fig: concept}-B) and the degree of support sought in users’ posts (\ie, \textit{v4} in \autoref{fig: concept}-B)---different proactive behavior types can be derived based on the attribute ranges of these variables (e.g., a behavior type characterized by high post frequency and low sought support degree).
    
    \item \textit{\textbf{Behavior events}} (\autoref{fig: concept}-C) comprehensively depicts a user's behavior within a time unit (\eg, within a day).
    Each behavior event includes a mental health status (\hspace{1mm}\typeiconm), a proactive behavior type (\hspace{1mm}\typeiconp), and a reactive behavior type (\hspace{1mm}\typeiconr). 
\end{itemize}
Over a long period of time (\eg, a year), a user’s behaviors form a well-structured behavior sequence (\autoref{fig: concept}-C), which is comprised of a list of behavior events by each time unit. 
We can further divide behavior sequences into a set of \textit{behavior stages} according to the temporal patterns of behavior events. 
\begin{itemize}[leftmargin=1em]
    \item \textit{Behavior stages} (\autoref{fig: concept}-D) are derived from behavior sequences to represent periods of relatively sustained behaviors~\cite{liu2017assessing}. 
    Notable shifts in user behaviors can indicate a potential change in their behavior stage.
    For example, consider a simplified scenario where each behavior event includes only a proactive behavior type and a reactive behavior type, each focusing on a single behavior variable (\textit{v2} in \hspace{0.8mm}\typeiconp, \textit{v7} in \typeiconr).
    Then, one behavior stage (\texttt{$stage_b$} in \autoref{fig: concept}-D) could be characterized by a continuous increase in post frequency (\textit{v2}) and a stable response number (\textit{v7}) over time.  
    Subsequently, the behavior stage (\texttt{$stage_c$}) might change to another stage with a continuous decrease in post frequency and a simultaneous increase in the response number over time.
    In our system, each behavior stage is defined by a specific range of mental health status, along with distinct value ranges for all proactive and reactive behavior variables.
\end{itemize}
The progression of behavior stages is referred to as \textit{stage progression}~\cite{liu2017assessing}, such as \texttt{$stage_b$-$stage_c$-$stage_b$-$stage_c$}. Finally, \textit{\textbf{stage patterns}} indicate commonly occurring individual stages or stage transitions, such as \texttt{$stage_b$-$stage_c$} (\autoref{fig: concept}-E).

\subsection{Problem Formulation}
Analyzing user behavior data from OMHCs is valuable for social scientists in uncovering group-level behavioral patterns related to mental health. Unlike clinical psychologists focused on individual diagnosis, social scientists aim to understand social dynamics and behavioral trends. This population-level insight aids in resource allocation, preventive strategies, and platform interventions, supporting clinical practice and public health initiatives.

\sidecomment{R2C1, R3C1}
\rui{We conducted formative studies with our experts, \Ea, \Eb, and \Ec.}
Early on, social scientists relied on self-reported survey data to estimate mental health status, which was limited by recall bias, non-responses, and poor temporal granularity. With the rise of social media, however, researchers can now infer mental health status from user-generated digital traces, such as posting patterns, content, and interactions.
Unlike surveys, these traces enable unobtrusive, large-scale, and real-time observations of behavior, allowing researchers to track changes over time. This shift provides a more dynamic, objective, and scalable approach to studying mental health in natural settings. Experts typically analyze user-generated content from OMHCs to examine behavior patterns across different mental health groups, aiming to inform intervention strategies.
A common approach is to segment behavior sequences into distinct stages~\cite{liu2017assessing, kushner2020bursts, peng2020mobile, ernala2021social}, such as high or low activity periods, and analyze how these stages change over time with users’ mental health trajectories. These analyses often require custom coding in statistical environments like R or Python to accommodate the complexity of longitudinal behavioral data.

Despite efforts, user behavior in OMHCs remains diverse, with actions like posting content and receiving responses. Jointly analyzing multiple behavior variables provides a more comprehensive view of users' mental health. However, traditional hypothesis-driven approaches, relying on predefined behavioral dimensions, struggle to capture the complexity of multivariate data. As behavior variables increase, so does the complexity of stage definitions, making interpretation harder and complicating the identification of patterns that affect mental health. Based on expert interviews, we identified two key design goals for a tailored visual analytics system: (\textbf{G1}) extracting behavior stages over time considering multiple variables, and (\textbf{G2}) identifying stage transition patterns linked to mental health outcomes.

\subsection{Requirement Analysis}\label{task}
Through extensive interviews and iterative feedback with our domain experts (\Ea, \Eb, \Ec), we identified five core tasks that align with the two primary goals of our system.
\begin{itemize}[leftmargin=1.8em]
    \item[\textbf{T1}] \textbf{Understand the multidimensional behavior variables}.
Our experts emphasized the difficulty of making sense of numerous behavior variables at once.
For example, \Ea mentioned, \textit{``When we look at post frequency, support-seeking, and response number together, it becomes too overwhelming to interpret them simultaneously.''}
To reduce this cognitive load, they recommended grouping the behavior variables into higher-level categories, such as user-initiated (e.g., posting) versus externally received (e.g., replies from others).
This serves as a necessary foundation for understanding various behavior stages (\textbf{G1}).

    \item[\textbf{T2}] \textbf{Distill and summarize behavior stages.}
Based on findings from our formative study, rather than analyzing behaviors at every time point, experts preferred identifying coherent behavior stages.
As \Ec noted, \textit{``We are not just interested in what someone does on a single day, but in stretches of time when their behavior is stable or changing in a meaningful way.''}
Therefore, our system should process these multivariate event sequences and extract distinct behavior stages, enabling experts to gain a more interpretable and higher-level understanding of users’ behavior trajectories (\textbf{G1}).

    \item[\textbf{T3}] \textbf{Summarize the behavior stage progression over time}.
Experts also highlighted the importance of tracking how users transition across stages over time and understanding the overall distribution of these transitions. 
For example, \Eb mentioned, \textit{``Understanding the progression of behavior stages across the group helps me get a holistic sense of which stages require more focused attention.''} Thus, the system should summarize stage transitions across the group to facilitate longitudinal analysis (\textbf{G2}).

    \item[\textbf{T4}] \textbf{Reveal frequent stage patterns and their impact on long-term mental health}.
The experts emphasized the needs to identify recurring stage patterns and link them to users’ long-term mental health outcomes. For example, \Ea noted, \textit{``There can be multiple stage patterns. I would like to directly examine the frequency of different stage patterns and assess how these patterns influence users' subsequent behaviors.''} By surfacing such stage patterns, experts can prioritize those most relevant for further examination and analysis (\textbf{G2}).

    \item[\textbf{T5}] \textbf{Reveal the context information of a focused \behaviorunit{} pattern}.
Once a potentially critical stage pattern is identified, experts require a comprehensive understanding of its full context. As \Ec emphasized, \textit{``It is interesting to learn what factors led up to a stage pattern and what happened after this pattern, so that we can assess whether an intervention is necessary.''} Therefore, our system should provide contextual information by integrating all behavior sequences within the group, offering a comprehensive understanding of the selected stage pattern (\textbf{G2}).
\end{itemize}{}

\section{Automatic Mining Pipeline}
\begin{figure*}[ht]
\centering
\includegraphics[width=1\linewidth]{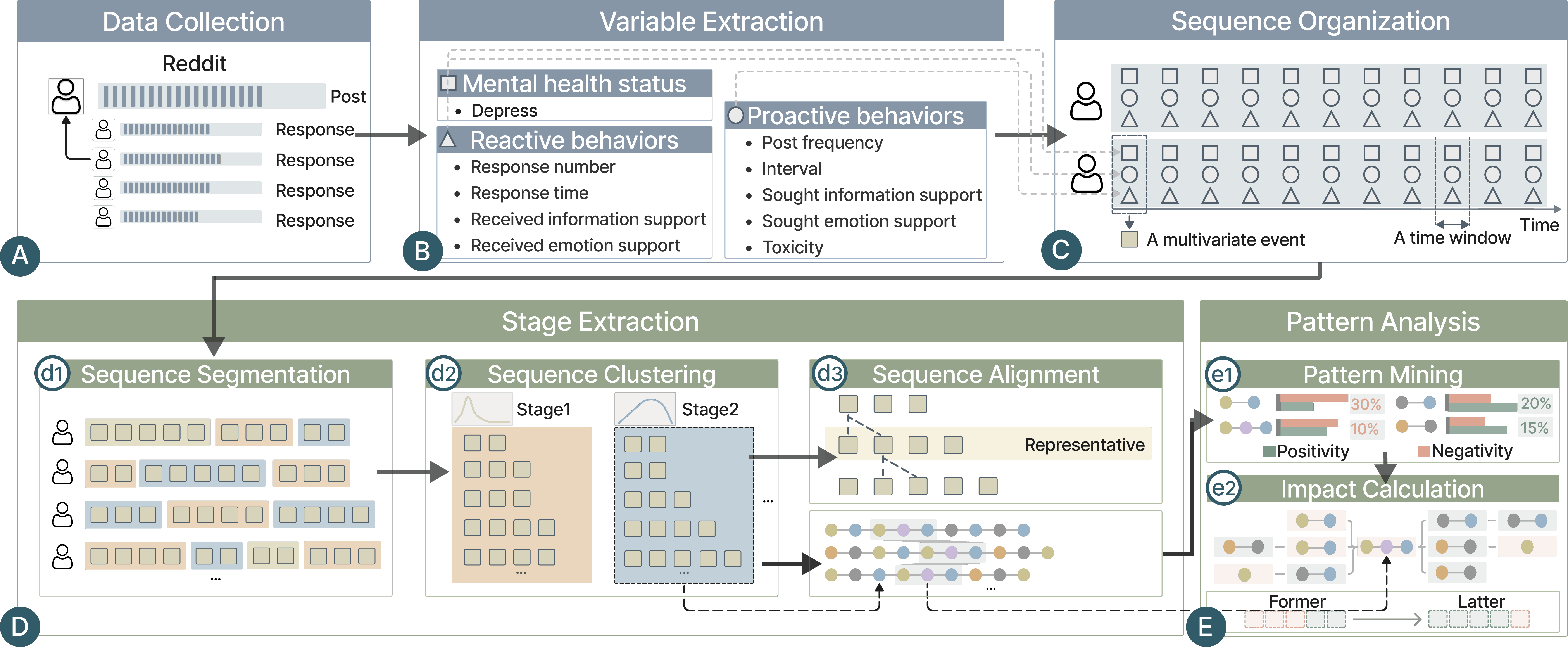}
\caption{The pipeline of the data processing and mining components. We first process the raw data into multivariate event sequences (A-C). Then the sequences will be segmented and clustered to form behavior stages (d1-d2). To present the temporal information within a behavior stage, we align the segmented sequences within a stage to obtain statistical information (d3). Additionally, frequent stage patterns are extracted (e1) through all the stage sequences. Meanwhile, the impact of each pattern on long-term mental health is calculated (e2).}
\label{fig: algorithm_workflow}
\vspace{-1em}
\end{figure*}

We developed an automatic mining pipeline to extract behavior variables from the collected data and construct multivariate event sequences. Subsequently, we extracted behavior stages and calculated the impact of frequent stage patterns. The pipeline is illustrated in \autoref{fig: algorithm_workflow}. 
To facilitate better reproducibility of our pipeline, we also provide detailed pseudocode that outlines the entire pipeline, which is included in the Appendix.

\subsection{Data Processing}\label{rawdata}
\subsubsection{\textbf{Data Collection}}\label{mh}
Reddit is a popular online forum with various communities (aka ``subreddits''), including some for mental health support.
We collected the user data from Reddit in 2020, which consists of users' posts along with the corresponding responses received and their timestamps using the PushShift API\footnote{https://github.com/pushshift/api} (\autoref{fig: algorithm_workflow}-A), allowing experts to explore the impact of user behaviors on mental health outcomes during COVID-19.
Based on the literature and discussion with experts, we categorize the subreddits into three categories: 1) Suicide support subreddit \textit{r/SuicideWatch} (\textbf{SW}), where community members present suicidal ideation, the late stage of depression; 2) Mental Health support subreddits (\textbf{MH}) where members could start to seek support for mental health issues (\ie, 15 commonly discussed mental health subreddits such as \textit{r/depression} and \textit{r/mentalh}~\cite{low2020natural}); and 3) all the other subreddits (\textbf{OT}) which show little relevance to mental health issues. 
Subreddits that do not belong to the \textbf{SW} subreddit or the \textbf{MH} subreddits~\cite{sharma2018mental} can be seen as \textbf{OT} subreddits.

To illustrate the behavior progression, we analyzed 1072 active users (posting at least twice a month) who participated in the MH subreddits during the first six months of the year and continued participation during the second half.
Based on their activity data in the second half year, we divided users into three transition groups: 1) \textbf{MH-SW} for those active in the \textbf{SW} subreddit, 2) \textbf{MH-MH} for those active in the \textbf{MH} subreddits but not in the \textbf{SW} subreddit, and 3) \textbf{MH-OT} for those active in other communities.

\subsubsection{\textbf{Variable Extraction}}\label{variable_extraction}
Based on the literature review~\cite{naslund2020social, kushner2020bursts, de2013predicting, 
peng2021effects, Saha_Sharma_2020, parker2010taking, rozo2016learning, chikersal2020understanding, 
almerekhi2022investigating} and expert input, we extracted behavior variables from the collected user data. 
We considered three kinds of information (\autoref{fig: algorithm_workflow}-B), including the mental health status of users, variables related to \textit{\activebehaviors{}}, and variables related to \textit{\passivebehaviors{}}, to construct \behaviorunits. 
The information encapsulates users' actions and acquisitions during a given period.

\textbf{Mental Health Status} reflects an underlying variable influencing user behaviors. We selected the degree of depression as a measure of mental health status, a commonly adopted metric in mental health research~\cite{de2013predicting, chikersal2020understanding}. 
We calculated the depression degree of each post through an ensemble algorithm~\cite{poswiata-perelkiewicz-2022-opi}, classifying posts into three levels: \textit{not depressed}, \textit{moderately depressed}, and \textit{severely depressed}, achieving an accuracy of 0.7 on the dev dataset and 0.66 on the test dataset.

\textbf{Proactive Behaviors} are initiated by users themselves~\cite{parker2010taking}. To quantify them, we extracted the following information:
\begin{itemize}[leftmargin=1em]

    \item \textit{Post frequency}: The number of posts~\cite{kushner2020bursts}.
    \item \textit{Interval:} The period of time during which users remain inactive, neither creating nor responding to posts~\cite{naslund2020social}.
    \item \textit{Sought information support}: The degree of asking for information to help understand a situation better, which can be obtained through the model proposed by Peng \etal~\cite{peng2021effects}. 
    \item \textit{Sought emotion support}: The degree of actively asking for comfort from others to relieve emotional distress, which can be obtained through the model proposed by Peng \etal~\cite{peng2021effects}. 
    \item \textit{Toxicity}: The degree of malicious content in the form of harassment, profanity, and cyberbullying that users take the initiative to read~\cite{almerekhi2022investigating}. We measured the toxicity degree of the posts that a user was exposed to by analyzing the posts on which they took the initiative to comment. Specifically, the toxicity degree can be calculated by a fine-tuned BERT model~\cite{davidson2017automated}, which is the commonly used toxicity prediction algorithm for social media content\footnote{https://huggingface.co/unitary/toxic-bert}. 
\end{itemize}

\textbf{Reactive Behaviors} are responses to user behaviors exhibited by others \cite{rozo2016learning}, illustrating how others’ actions impact the user. We extracted the following variables:
\begin{itemize}[leftmargin=1em]
    \item \textit{Response number}: The number of responses toward users' posts~\cite{Saha_Sharma_2020}.
    \item \textit{Response time}: The average time interval between the user-initiated posts and the corresponding first response~\cite{Saha_Sharma_2020}.
    \item \textit{Received information support}: The degree of the received information support can be obtained through the model proposed by Peng \etal~\cite{peng2021effects}. 
    \item \textit{Received emotion support}: The degree of the received emotion support can be obtained through the model proposed by Peng \etal~\cite{peng2021effects}. 
\end{itemize}

\subsubsection{\textbf{Sequence Organization}}\label{so}
We organized the data into behavior sequences (\autoref{fig: algorithm_workflow}-C) as follows:

\textbf{Behavior variable extraction}. 
We calculated the average value of each behavior variable within a time window. To balance between the excessive smoothing of a month-long window and the noise of a week-long window, we selected a two-week window for our study. This length can be adjusted depending on analysis needs. Longer windows provide smoother results but may miss sudden behavior changes (e.g., escalating suicidal tendencies), while shorter windows (e.g., two to three days) better capture such emergent behaviors. At each time unit, we obtained the mental health status along with the proactive and reactive behavior variables.

\textbf{Behavior type distillation}. 
To facilitate expert analysis, we clustered proactive and reactive behaviors separately, allowing behavior variables to be examined from an aggregated perspective. We employed the K-Means algorithm to derive distinct behavior types within each category, providing a structured representation of user behavioral patterns.

\textbf{Behavior event formation}. 
We combined the depression degree, \activebehavior{} type, and \passivebehavior{} type at each time unit to form a multivariate behavior event. Consequently, each behavior sequence is represented as a temporally ordered series of such multivariate events, capturing the dynamic evolution of user behaviors over time.

\subsection{Mining Algorithms} 
We describe our methods for extracting common behavior
stages (\autoref{fig: algorithm_workflow}-D) and identifying stage patterns that impact subsequent behaviors (\autoref{fig: algorithm_workflow}-E).

\subsubsection{\textbf{Stage Extraction}}\label{stage_extraction}

 We introduced the following three steps to extract and summarize various behavior stages from all behavior sequences:

\textbf{Sequence segmentation}. 
We segmented users’ behavior sequences by detecting notable changes using time series segmentation algorithms (\autoref{fig: algorithm_workflow}-d1). 
To identify the most suitable algorithm, we first constructed a small labeled dataset and evaluated several widely used unsupervised segmentation methods. Supervised approaches were excluded due to the limited size of our labeled data.
Among the four methods (\ie, HMM~\cite{Sean1996Hidden}, GGS~\cite{hallac2016ggs}, TIRE~\cite{de2021change}, AutoPlait~\cite{Matsubara2014AutoPlait}), GGS achieved the best performance, with an accuracy of 0.725.
The details can be seen in the appendix.
GGS detects shifts in behavior, modeling each sub-sequence as an independent sample from a Gaussian distribution~\cite{hallac2016ggs}, thus capturing variations in user behavior over time. 
After applying GGS, each behavior sequence is divided into several varied-length sub-sequences, each representing a unique behavior stage distinct from preceding and succeeding periods.

\textbf{Sequence clustering}.
We aimed to summarize common behavior stages among all users by clustering similar sub-sequences. 
Direct clustering based on the Euclidean distances between sub-sequences may not adequately capture the characteristics of trends. 
Therefore, we benefited from the GGS used in the previous step. 
We utilized the parameters of the Gaussian distribution for each sub-sequence as measures of similarity.
Specifically, we obtained the mean vector ($\mu^{1 \times n}$) and the covariance matrix ($\Sigma^{n \times n}$) of each sub-sequence.
We constructed a vector $v^{1 \times n}$ for clustering the sub-sequences, and $v = (1-\lambda) \times \mu^{'} + \lambda \times \Sigma^{'}$, where $\mu^{'}$ and $\Sigma^{'}$ represent the normalized results for $\mu$ and $\Sigma$, respectively, and  $\lambda$ is used to control the impact of covariance. In this experiment, we set the value of $\lambda$ to 0.1 (see the Appendix for the detailed reasons).
Then we used the K-means algorithm to cluster all the sub-sequences based on the calculated vector $v^{1 \times n}$, mainly as K-means is well-suited for this task due to its ability to efficiently handle the spherical clustering structure of the data.
\sidecomment{R3C4}
\rui{Specifically, the $\mu^{'}$ vectors are scale-balanced across dimensions (\ie, values across dimensions are comparable, so no single dimension dominates distance calculations), which ideally leads to similar sub-sequences clustering together in a spherical structure. Additionally, each $\Sigma^{'}$ is also balanced across dimensions, and in high-dimensional space, the distribution around each $\mu^{'}$ ideally forms a spherical shape. These factors make K-means the preferred clustering method. 
In contrast, density-based clustering (e.g., DBSCAN) may merge regions through density reachability, which can sometimes connect clusters that are close in dense areas but separated in feature space, thus obscuring meaningful cluster boundaries. Hierarchical clustering is sensitive to small perturbations; even a few noisy subsequences can change the dendrogram and yield unstable clusters. Therefore, considering K-means' alignment with the clustering requirements, its high performance, and easy settings, we chose K-means over methods like DBSCAN or hierarchical clustering.
Experts can adjust the number of clusters through our interface after examining the clustering results.
We have provided several hints to help them make better choices for clustering (\ie, the silhouette score of the clustering and the differences between those clusters).
A silhouette score provides an overall assessment of the clustering results, with a higher score indicating better cohesion and separation. However, since experts may prefer a more intuitive understanding of the differences between clusters, we also provide cosine differences to assess their distinctness. Experts can combine the silhouette score, cosine differences, and the distribution of cluster sizes to adjust the number of clusters.
}
The summarization of each cluster can be seen as a \behaviorunit{} (\autoref{fig: algorithm_workflow}-d2).
Specifically, \autoref{fig: teaser}-B visualizes the results of clustering—six distinct behavior stages. For instance, \autoref{fig: teaser}-\texttt{stage2} (the third one in \autoref{fig: teaser}-B due to the dragging operation) indicates the cluster of sub-sequences with high posting frequency and numerous received responses, as indicated by the higher proportions of p1, p4, and r3. This is clearly distinguishable from \autoref{fig: teaser}-\texttt{stage1}, where the lower proportion of \texttt{r3} indicates fewer received responses (the meaning of \texttt{r3} can be checked in \autoref{fig: teaser}-A).

\label{sequence_alignment}
\textbf{Sequence alignment}. To represent each behavior stage, we aggregated all sub-sequences within a cluster. To capture temporal patterns, we aligned all sub-sequences to a fixed length (set to 5 in our experiment) and quantified behaviors at each time unit (\autoref{fig: algorithm_workflow}-d3). This length was chosen to balance capturing sufficient temporal information while avoiding excessive complexity.
To complete the alignment, we first identified representative sub-sequences by filtering sub-sequences of length 5 and applying hierarchical clustering, which is more effective at handling outliers during the clustering process. Then, we selected 3 representatives per cluster based on clustering evaluation metrics (Calinski-Harabasz and Silhouette scores). Next, we leveraged Dynamic Time Warping (DTW) algorithms to align all sub-sequences within a stage to the closest representative. Specifically, the Weighted Derivative Dynamic Time Warping (WDDTW) approach was used due to its superior performance in complex scenarios compared to other DTW methods~\cite{rajput2024evaluatingdtwmeasuressynthesis}. 
Finally, we obtained fixed-length sequences, allowing us to summarize behaviors at each time unit and discern temporal trends. 
The aggregated statistical information of aligned sub-sequences constitutes the meaning of the corresponding \behaviorunit.

\subsubsection{\textbf{Pattern Analysis}}\label{PA}
\begin{figure}[ht]
\centering
\includegraphics[width=\linewidth]{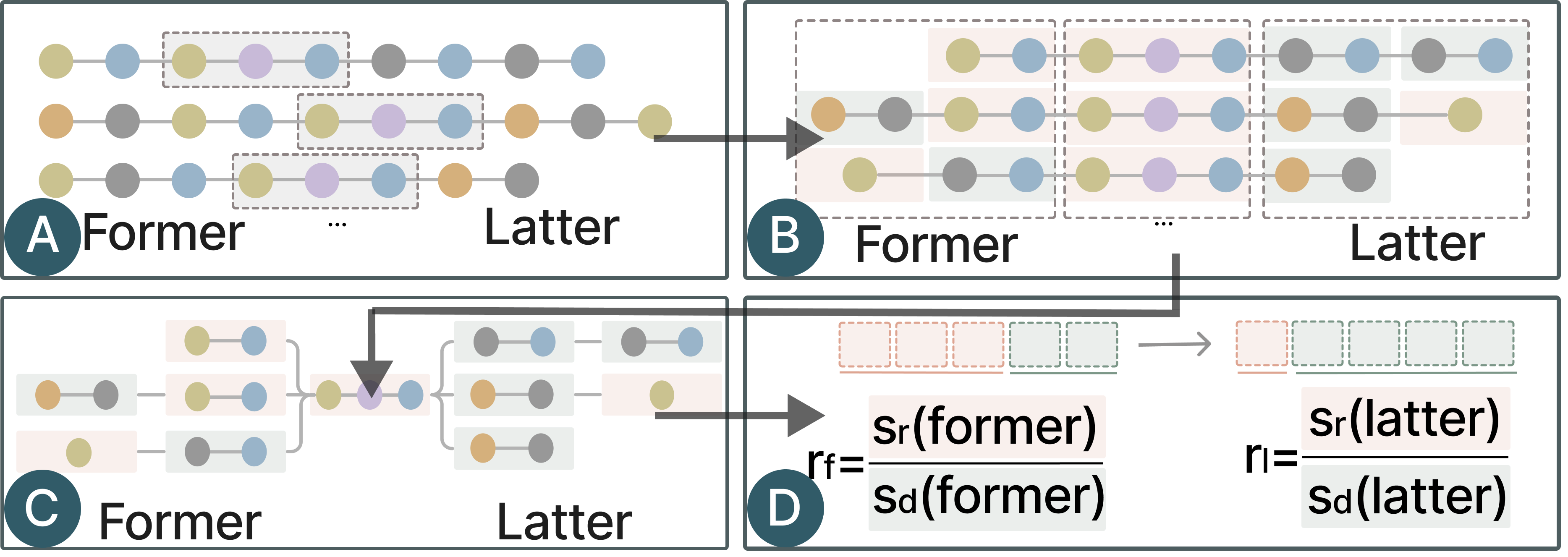}
\caption{Impact calculation. (A) Detect the selected pattern and divide the whole sequence into two sets. (B) Decompose all the sequences into frequent patterns. (C-D) Calculate the overall positivity or negativity in the two sets.}
\label{fig: impact}
\end{figure}

Experts aim to identify key stage patterns that may impact mental health by triggering lasting behavior changes. To assess their influence, we compare behaviors before and after a specific stage pattern, establishing relations between behavior changes and the pattern.
We first defined the positivity or negativity of a pattern based on the distinct transition groups classified in \autoref{mh}. Specifically, the frequency of a pattern can indicate its relevance with a group~\cite{sparck1988statistical}.
For example, the \textbf{MH-SW} group represents users whose condition deteriorated (deterioration group), while the \textbf{MH-OT} group serves as a comparison group (recovery group). 
Behaviors occurring more frequently in the deterioration group but not in the recovery group are considered more negative. 
Based on the extracted \behaviorunits{}, we measure the impact of each \mbox{\unit~(\autoref{fig: algorithm_workflow}-E) as follows:}

\begin{figure*}[ht]
\centering
\includegraphics[width=1\linewidth]{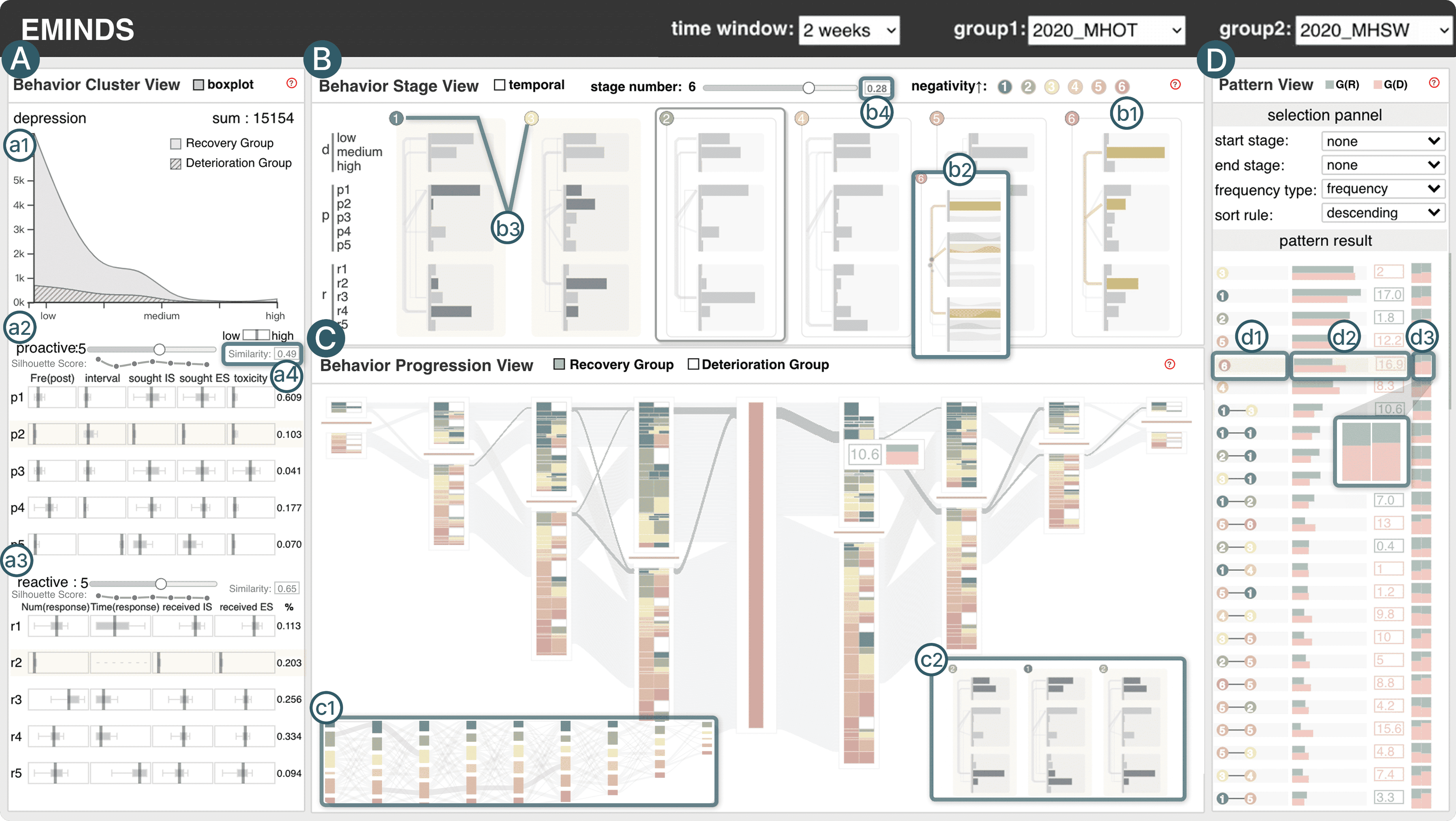}
\caption{The interface, featuring: (A) Behavior Cluster View, (B) Behavior Stage View, (C) Behavior Progression View, and (D) Pattern View.}
\label{fig: teaser}
\end{figure*}

\textbf{Pattern mining}.\label{pm}
We extracted frequent stage patterns using pattern mining methods.
Due to the short length of behavior stage sequences, we employed the exhaustive method to count all potential patterns directly. The frequent patterns were filtered using a threshold, with a default frequency of 0.1, considering both the occurrence frequency and pattern length diversity.
We then obtained the positivity or negativity of a stage pattern based on its frequency in different groups.
In this case, if a pattern appears much more often in the group of recovery than in the group of deterioration, it means that this pattern may represent more positive (\autoref{fig: algorithm_workflow}-e1, the left column).
It can be represented by $w(p) = f_{\text{r}}(p)-f_{\text{d}}(p)$, where $f_{\text{r}}(p)$ and $f_{\text{d}}(p)$ represent the frequency of this pattern in the recovery group and deterioration group, respectively. 
Therefore, the sign of $w(p)$ indicates whether it is positive or negative, while the magnitude of $w(p)$ represents the degree of positivity or negativity.
As an illustration, the frequent stage pattern \texttt{stage1-stage5} (shown at the bottom of \autoref{fig: teaser}-d3) is observed 3.3\% more frequently in the deterioration group, suggesting slight negativity.

\textbf{Impact calculation}. To identify patterns that might have a long-term impact on mental health, we need to measure how a pattern influences behaviors after its occurrence. 
We borrowed the concept of pre-post event analysis commonly found in social science~\cite{Christina2018Changing}, whose core idea is to measure the changes in observational variables before and after the occurrence of an event.
In our project, the changes in the positivity or negativity of behaviors are the observational variables social scientists are concerned with.
To compute the impact of a stage pattern, we first divide all stage sequences containing this pattern into two parts based on its position (\autoref{fig: impact}-A).
The former set represents the \behaviorunit{} sequences before this stage pattern appears, and the latter set represents the \behaviorunit{} sequences after the stage pattern appears. 
Then, we summed up the positivity or negativity degree for all the stage patterns in the former sequence set and the latter sequence set (\autoref{fig: impact}-C), respectively. 
Therefore, we obtained the overall positivity ($s_{\text{r}}$) or negativity ($s_{\text{d}}$) degree of both the two set (\autoref{fig: impact}-D).
Next, we calculated $r_{\text{f}} = \frac{s_{\text{r}}(\text{former})}{s_{\text{d}}(\text{former})}$, and $r_{\text{l}} = \frac{s_{\text{r}}(\text{latter})}{s_{\text{d}}(\text{latter})}$.
One thing to note is that the length of the former or latter sequences may affect results. When the lengths are too short, the calculated ratio can be overly influenced by a specific pattern. To avoid this situation, we discarded sequences that were shorter than the average length of the former and latter sequences.
We leveraged the difference between $r_{\text{f}}$ and $r_{\text{l}}$ to represent the impact of this selected pattern: $d_{\text{i}} = r_{\text{f}} - r_{\text{l}}$.
The larger the absolute value of $d_{\text{i}}$ indicates the greater difference between the former sequences and the latter sequences of the given pattern (\autoref{fig: algorithm_workflow}-e2).
\section{Visual Design}
\systemname provides four coordinated views.
The workflow begins with experts examining the distribution of behavior variables and summarizing behavior types through clustering in the \behaviourclusterview (\autoref{fig: teaser}-A).
Subsequently, they explore the \unit{} visualizations in the \progressionstageview (B) and obtain an overview of the progression of the selected groups in the \progressionview (c6).
Next, experts will proceed to explore the \patternview (D) to select a stage pattern they are interested in.
Additionally, a \newSankey in which all the sequences are aligned based on the selected stage pattern (c1) is displayed to help experts comprehend the context of the stage progression. 
The system implementation details can be seen in the Appendix.

\subsection{Behavior Cluster View}
\sidecomment{R3C2}
\rui{This view (\autoref{fig: teaser}-A) is aimed to help understand the multidimentional behavior variables (\textbf{T1}), including the \depressiondegree{} (a1), the proactive behaviors (a2), and the reactive behaviors (a3).
Based on the results of the formative study, proactive and reactive behaviors are presented using behavior type as the basic unit. This approach helps users better interpret common combinations of behavior variables.
} 
 
First, the depression distribution is visualized through area charts, where the horizontal axis represents the AI-predicted depression outcomes (less than 1 indicates not depressed, 1-2 indicates moderately depressed, and 2-3 indicates severely depressed), while the vertical axis represents the number of posts. 
Additionally, we display the total number of posts in the top right corner of this view.
Then, we adopted a matrix design to display behavior types.
\sidecomment{R3C2}
\rui{This matrix not only helps users quickly understand the behavior variables that constitute each behavior type, but also allows experts to effectively verify and compare differences in behavior variables across different behavior types. By facilitating rapid comparisons, it enables social scientists to better understand new behavior types through the lens of previously explored ones. This approach further alleviates the cognitive load associated with interpreting multidimensional behavior variables.}
Specifically, in the matrix, each row represents a behavior type (\eg, \texttt{p1} and \texttt{r1} in (\autoref{fig: teaser}-A), which consists of multiple behavior variables (\eg, \texttt{Fre(post)}, \texttt{interval}, \texttt{sought IS}) with different values. 
Matrix cells display box plots representing the distribution of values for each behavior variable.
For example, \texttt{r1} indicates a large number of responses and high levels of received information support, while \texttt{r2} reflects few responses and lower support levels.

\textit{Interaction:}
\sidecomment{R3C4, R3C2}
\rui{Experts can control the number of clusters for active behaviors and passive behaviors using the slider. The silhouette score provides an overall overview to assist in their selection. 
When hovering over a corresponding point on the line chart, the specific silhouette score value will be displayed.
Additionally, since social scientists typically prefer a moderate number of evenly distributed and sufficiently distinct clusters, the system also visualizes both the similarity score (\autoref{fig: teaser}-a4) that reflects the degree of overlap among clusters and the proportion of data samples in each cluster.} These visual cues help experts assess the balance and separation of clusters when determining an appropriate clustering solution.

\subsection{Behavior Stage View}\label{stage}

\begin{figure}[ht]
\centering
\includegraphics[width=\linewidth]{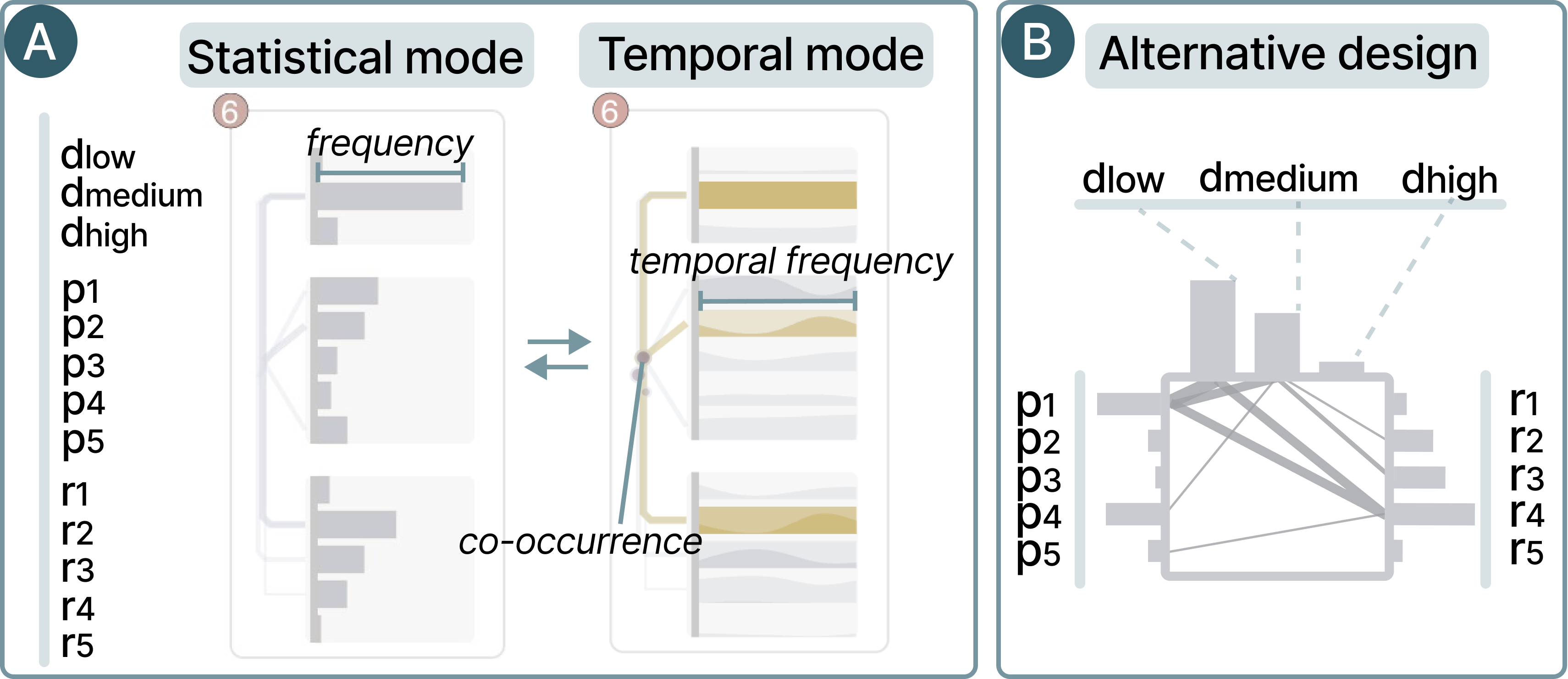}   
\caption{(A) Two modes of the behavior stage visualization (\ie, statistical and temporal). (B) The alternative design.}
\label{fig: stage}
\end{figure}

\sidecomment{R3C2}
\rui{This view (\autoref{fig: teaser}-B) visualizes the \behaviorunits{} automatically extracted and summarized through our pipeline, directly supporting experts in the task of distilling and summarizing coherent behavior stages (\textbf{T2}). Experts highlighted that analyzing behaviors at every single time point is neither meaningful nor scalable, since interpretation requires considering how behaviors evolve within broader temporal contexts. To address this, our design aggregates behavior events into higher-level stages that capture stable or meaningful transitions. To further support interpretation, we provide two complementary modes: the statistical mode, which condenses occurrence and co-occurrence frequencies of behavior types to offer a concise overview of a behavior stage and the temporal mode, which reveals how these behavior types unfold and interact over time in a behavior stage. Together, the two modes let experts shift between abstraction levels—quickly spotting an overview and, when needed, examining detailed temporal dynamics—thus reducing the cognitive load of interpreting complex behavior stages.}
Furthermore, behavior stages are ordered by their positivity, where \texttt{stage1} indicates the highest positivity.
Given that experts typically avoid considering an excessive number of stages to prevent overlapping or similar stages, we employ a color gradient ranging from deep green to orange, corresponding to the positivity degree of the stages. This color scheme allows experts to distinguish between stages without overwhelming them with too many colors.
\\
\indent \textbf{Statistical mode}. To give experts an overview of a behavior stage, we initially show the occurrence frequency of each behavior type within a stage (\autoref{fig: stage}-A). 
The \depressiondegree, the \activebehavior{} types, and the \passivebehavior{} types are represented in distinct regions, respectively, aligned with the order in the \behaviourclusterview.
This can help experts quickly recall the semantic meaning of each behavior type.
The length of the bar represents the frequency of the variable occurring at this stage.
For instance, \texttt{p1} and \texttt{p2} are the most frequent proactive behavior types in \texttt{stage6}, while \texttt{r2} and \texttt{r3} are the most frequent reactive behavior types.
Then, we utilize lines to display the co-occurrence frequency between corresponding behavior types, calculated by counting their occurrences in the same behavior event.
The thicker the line, the higher the degree of co-occurrence.
Only frequent co-occurrences are shown, which can reduce experts' overload.
For example, when the depression is middle, \texttt{p2} and \texttt{r2} will always occur together, which means the three have a high interaction (\autoref{fig: stage}-A).
To better display those lines, we set the line layout by calculating the distance between the connected behavior types: when the associated behavior types are further apart, the lines will also be positioned farther from the bars.

\sidecomment{R3C2}
\indent \textbf{Temporal mode}. \rui{The temporal mode (\autoref{fig: stage}-A) is designed to address the need for understanding dynamic changes within a stage, which cannot be captured by static statistics alone. By transforming overall frequencies into aligned temporal sequences (\autoref{sequence_alignment}), the temporal mode allows experts to inspect how different behavior types rise, peak, and decline over the course of a stage. Each behavior type’s temporal trajectory is presented in the same spatial order and semantic grouping as in the statistical mode, preserving mental mappings and facilitating smooth transitions between overview and temporal inspection. Experts can identify patterns such as proactive behaviors triggering reactive responses or shifts in depression-related signals over time, which are critical for understanding potential causal or sequential relationships.}

\textit{Alternative:} \rui{To support the task of understanding the internal relationships among behavior types within a stage (\textbf{T2}), we initially adopted a square ring design (\autoref{fig: stage}-B), placing behavior types on the outer rim and encoding interactions inside. 
This design aimed to provide a compact presentation of co-occurrences and facilitate inspection of relationships within a stage.
While compact, the layout made it difficult for users to quickly recall the semantic meaning of each behavior type due to its inconsistent ordering and misalignment with the sequence in the Behavior Stage View.
Furthermore, the layout made it difficult to compare different behavior stages and did not support inspection of the temporal dynamics within a stage.
Therefore, we adopted the ordering used in the Behavior Cluster View and visualized interactions on the left using connecting lines.
We also considered showing statistical and temporal information simultaneously, but this overwhelmed experts with too much visual detail.
To address these challenges, we adopted a two-mode design: the statistical mode provides a concise overview of occurrence and co-occurrence frequencies, while the temporal mode offers detailed within-stage dynamics as additional context.}

\textit{Interaction:}
First, experts can determine the stage number using a method similar to that employed in the Behavior Cluster View (\autoref{fig: teaser}-b4).
Second, the corresponding behavior types will be highlighted when experts hover over an interaction linkage. 
In addition, when hovering over a behavior type, any other behavior types that have interacted with it will be highlighted (\autoref{fig: stage}-A). 
To help experts better interpret the meaning of behavior stages, we also coordinate this view with the Behavior Cluster View.
Specifically, when a behavior type in the \progressionstageview is highlighted (\autoref{fig: teaser}-b1), the corresponding behavior type will be simultaneously highlighted in the \behaviourclusterview (\autoref{fig: teaser}-a2, a3).
Third, experts can click the behavior stages to compare them. After clicking, behavior types with notable differences across the selected behavior stages will be highlighted (\autoref{fig: teaser}-b3). 
This is done by calculating the variance of the corresponding behavior type frequencies for each stage.
Finally, experts can drag each stage visualization and re-order them for better comparison.

\begin{figure*}[t]
\centering
\includegraphics[width=\linewidth]{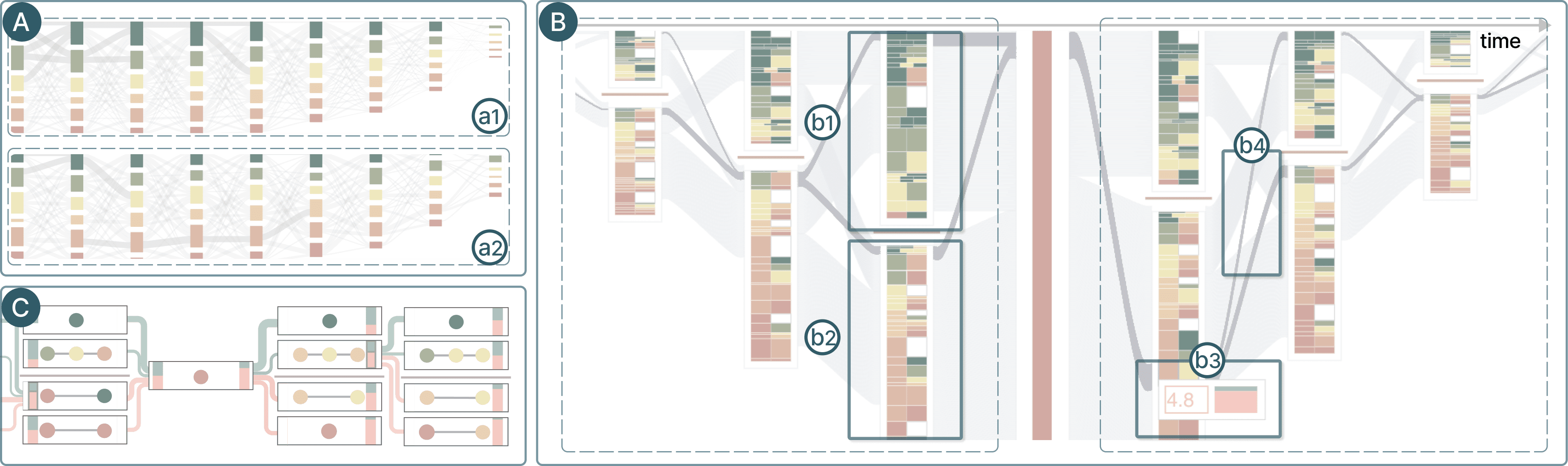}   
\caption{(A) Sankey diagrams to provide
an overview of the stage progression of the recovery group or the deterioration group. (B) A pattern-centric Sankey diagram shows the evolution of a specific stage pattern over time. (C) The alternative design.}
\label{fig: progression}
\vspace{-1.5em}
\end{figure*}

\subsection{Behavior Progression View}\label{progression}
\sidecomment{R3C2}
\rui{This view (\autoref{fig: teaser}-C) provides a comprehensive overview of the progression of behavior stages, addressing two key tasks identified in our formative study: summarizing stage transitions over time (\textbf{T3}) and revealing the contextual information of a focused \behaviorunit{} pattern (\textbf{T5}).}
Therefore, two kinds of diagrams are displayed to help experts explore the progression based either on groups (\autoref{fig: progression}-A) or assigned \behaviorunit{} patterns (\autoref{fig: progression}-B).

First, we leverage a Sankey diagram (\autoref{fig: teaser}-c2) to provide an overview of the stage progression. 
Experts can examine the stage progression of the recovery group and the deterioration group, respectively (\autoref{fig: progression}-A).
The Sankey diagram consists of nodes with distinct colors that represent different stages, corresponding to the \units{} exhibited in the \progressionstageview.
The height of each node indicates the frequency of \units{} occurring at that time unit.
The flow between the two adjacent nodes indicates the transition between \units.
\rui{Sankey diagrams are particularly suitable here because they naturally represent many-to-many transitions, allowing experts to track how multiple stages evolve into multiple subsequent stages over time. To reduce cognitive load, we highlight only the top ten highest-frequency flows, emphasizing the most important transitions. Color coding aligns with the Progression Stage View, supporting rapid recognition and consistent mapping.}

\sidecomment{R3C2}
\rui{When experts determine a specific stage pattern in the Pattern View (\autoref{fig: teaser}-D), the Sankey diagram is transformed into a \newSankey (\autoref{fig: progression}-B) (\textbf{T5}), revealing the full context of the selected pattern. 
This can help experts understand what factors might lead up to a stage pattern and what happened afterward.
To achieve this, we align all behavior sequences based on the selected pattern, separating them into preceding and following segments. This alignment allows experts to analyze behavior changes before and after the focal pattern, providing a clear temporal context for their interpretation.
Additionally, stage patterns at each time unit are aggregated and organized by positivity: patterns more frequent in the recovery group are placed at the top (\autoref{fig: progression}-b1), while negative patterns are grouped at the bottom (b2).
Node height encodes pattern frequency, maintaining consistency with the original Sankey diagram.}

\textit{Alternative:} 
\rui{To support experts in understanding the context of a selected stage pattern—specifically, what behaviors led up to it and what followed—we initially designed a layout (\autoref{fig: progression}-C) that displayed each type of stage pattern in the preceding and following sequences. The idea was to let experts trace the full temporal context of a pattern across all sequences.
However, because each node had a fixed height, patterns farther from the center caused the layout to expand rapidly, creating visual clutter.
More importantly, this design made it difficult for experts to distinguish the relative importance of different behavior stage patterns, such as identifying which ones occurred more frequently.
Using this Sankey-based design not only maintains consistency with the original representation (both employ Sankey diagrams), which reduces users’ mental switching cost, but also offers a more compact visual layout. This compactness enables experts to more easily observe overall temporal trends while still distinguishing the most important stage patterns as context for deeper analysis.
}

\textit{Interaction:} When hovering over a specific pattern, its positive or negative degree and its impact on long-term mental health will be shown (b3).
We also illustrate the transition of the positive and negative nodes through flow (b4).
Additionally, hovering over a stage pattern in the \progressionview displays the corresponding stage visualizations in a tooltip (\autoref{fig: teaser}-c2), aiding experts in grasping the stage transition information.

\subsection{Pattern View}
\sidecomment{R3C2}
\rui{This view (\autoref{fig: teaser}-D) presents frequent stage patterns within each group (\textbf{T4}).
Experts stressed the importance of identifying recurring stage patterns and linking them to outcomes. To meet this need, we list frequent patterns along with their frequencies in different groups and visualize their impact on subsequent behaviors, enabling experts to quickly prioritize the most influential ones.}
The impact, which is measured as the difference in sequences before and after a stage pattern is also shown in a glyph. Therefore, experts can swiftly focus on the important stage patterns. 
In the pattern result panel, we created a list to display information about frequent stage patterns.
Each row in the list represents a unique frequent \behaviorunit{} pattern.
The first element (\autoref{fig: teaser}-d1) in each row is the \unit{} order within the pattern.
The second element (d2) displays the frequency of the pattern occurring in two different groups, each represented by a distinct color. 
In addition, the concrete value of the frequency difference between the two groups is shown below.
The final element (d3) illustrates the changes in the positive and negative degrees of behavior sequences before and after this stage pattern.
\sidecomment{R3C2}
\rui{This design integrates both behavior prevalence and impact into a compact representation, enabling experts to quickly identify patterns of interest without cognitive overload. The layout also supports direct comparison between stage patterns, facilitating prioritization for subsequent analysis.}

\section{Case Study}\label{case}
We invited our expert \Ea introduced in \autoref{expert} to use our system and provide his exploration insights.
First, we guided him through the system step-by-step, after which he spent about 30 minutes familiarizing himself through free exploration. He then formally analyzed online user behaviors while thinking aloud, allowing us to document his process and findings over an hour.
Finally, we confirmed his findings with him and summarized some interesting insights into two cases to showcase our system's capabilities.

\subsection{Case I: Residual Vulnerability in Depression: Why Continued Support Matters}
In this case study, \Ea aimed to identify behaviors linked to long-term mental health deterioration and uncover insights overlooked by previous research. These findings could inform future research and support tasks like detecting mental health issues and designing interventions.
Specifically, he analyzed the user behaviors collected from Reddit in 2020, examining the \textbf{MH-SW} group and comparing it with the \textbf{MH-OT} group.
Firstly, \Ea determined the cluster number in the \behaviourclusterview by examining the cluster results of proactive behaviors. He found that when the value was set to 5, the similarity value and the distribution of clusters appeared more balanced (\autoref{fig: teaser}-a4). 
Similarly, he also set the cluster number for \passivebehaviors{} to 5.

\textbf{What are the common stage transitions?} Next, he explored \behaviorunits{} in the \progressionstageview{} (\autoref{fig: teaser}-B) and set the stage number to 6 (\autoref{fig: teaser}-b4), following a similar approach as the cluster number selection.
He hoped to find what kind of stage transition occurs frequently.
\textit{``We need to pay attention to common stage transitions. If a common transition can harm one's long-term mental health status, we need to be more attentive.''}
Then he observed that the transition between \texttt{stage1} and \texttt{stage2} was very common in the \progressionview{} since the flow between the top two stages was stronger (\autoref{fig: progression}-a1, a2) (\textbf{T3}).
This prompted him to investigate the potential impact of stage patterns involving these two stages on the user behavior trajectory.

\begin{figure}[h]
\centering
\includegraphics[width=\linewidth]{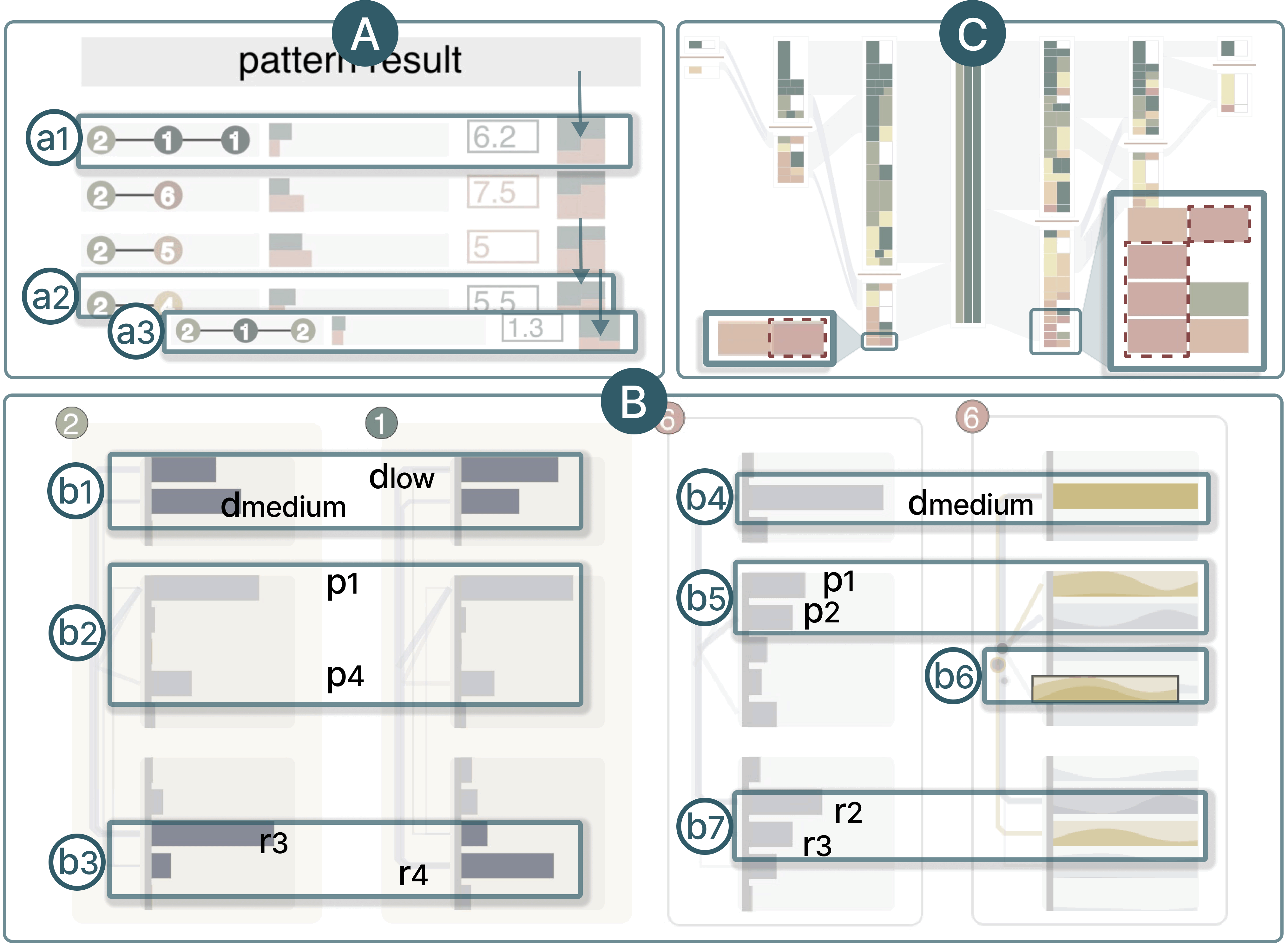}
\caption{Case I. (A) Determine a specific stage pattern. (a1) The pattern \texttt{stage2-stage1-stage1} led to a negative increase in the subsequent behaviors since the red bar went up a lot. (B) Understand the meanings of the behavior stages. (b1-b3) For example, from \texttt{stage2} to \texttt{stage1}, the depression decreased, but users still primarily exhibited \texttt{p1} and \texttt{p4}, indicating active seeking of social support, while reactive behaviors shifted from \texttt{r3} to \texttt{r4}, suggesting a decrease in received responses. (C) Check the detailed stage sequences. The expert found the frequency of \texttt{stage6} increased.}
\label{fig: case1}
\vspace{-1em}
\end{figure}

\textbf{Whether the desired patterns have a negative impact?}
To examine whether any stage patterns including \texttt{stage1} and \texttt{stage2} will have a negative impact on users' mental health status, \Ea started to explore the \patternview (\textbf{T4}).
He adjusted the widgets in the \patternview to find the patterns that included \texttt{stage1} and \texttt{stage2}.
Then, he sorted the patterns based on their negative impact.
Surprisingly, he observed a pattern \texttt{stage2-stage1-stage1} (\autoref{fig: case1}-a1) which would lead to a notable increase in the degree of negativity based on the glyph illustrating the impact change.

\textbf{What is the implication of this pattern?} \Ea then analyzed the pattern \texttt{stage2-stage1-stage1}.
Furthermore, \Ea hoped to understand the semantic meaning of this pattern (\textbf{T2}).
He first dragged the glyph of the \texttt{stage2} before the \texttt{stage1} (\autoref{fig: case1}-B) to align with the pattern order.
Through comparing the bar charts in the \progressionstageview, he found the differences between the two stages.
First, lower depression levels appear more frequently in \texttt{stage1}, as illustrated by a longer bar.
\textit{``There is a slight reduction in the depression degree from \texttt{stage2} to \texttt{stage1} {\rm (\autoref{fig: case1}-b1)}. It makes me wonder why the subsequent behaviors become more negative.''}
He examined proactive and reactive behaviors (b2, b3) and found that users were still actively seeking social support through posting on Reddit. The ratio of \texttt{p1} indicated normal posting frequency, followed by a high posting frequency in \texttt{p4} (b2). Comparing the reactive behaviors, he noticed an obvious difference in the number of received responses between the two stages (b3), with \texttt{stage2} dominated by \texttt{r3} and \texttt{stage1} dominated by \texttt{r4} (\textbf{T1}). Finally, \Ea concluded that the stage pattern indicated users experiencing a slight reduction in depression but still actively seeking social support, while the number of received responses decreased.

\textbf{What kind of behavior stages will follow?} 
After understanding this stage pattern, \Ea felt curious about how this pattern might influence subsequent behavior.
Therefore, \Ea tried to identify what kind of negative behaviors increased after the pattern \texttt{stage2-stage1-stage1} occurred (\textbf{T5}).
He observed the progression and found that the number of \texttt{stage6} increased after this pattern, as illustrated by the increased height of the nodes including \texttt{stage6} (\autoref{fig: case1}-C.
He clicked and checked the temporal information of \texttt{stage6} (\autoref{fig: case1}-B).
From the stage visualization, the depression of users was a little high (b4). 
Furthermore, the frequency of \texttt{p2} was increasing in this stage, as indicated by the temporal trend in the stage visualization, while \texttt{p1} was declining (b5).
Considering that the level of seeking social support for \texttt{p2} is lower than that of \texttt{p1}, \Ea understood that the frequency of seeking social support on Reddit was decreasing.
Similarly, \Ea also found that the number of responses and the amount of received social support were decreasing in \texttt{stage6} (b7).
\textit{``Users may be less motivated to engage with a community when they receive few responses. Surprisingly, even if their depression levels are high or increasing, it is unlikely that they would leave the community.''}

\textbf{Rethinking response number}.
\Ea initially believed that the number of responses would not impact users’ mental health according to his previous knowledge~\cite{Saha_Sharma_2020}.
However, using our system, \Ea discovered that the number of responses did play an important role in influencing users' mental health in some special periods. For instance, users may be more sensitive to the response number when relying on others' responses to aid their recovery from depression.
He remarked, \textit{``The different conclusion from the previous study may be due to the system enabling more detailed analysis, such as studying the impact of response number combining exploring multiple proactive behaviors and the dynamic changes in depression.''}



\subsection{Case II: Fluctuations in User Behaviors: Signals of Potential Mental Health Deterioration}
Continuing his exploration of the 2020 user behavior data, \Ea observed an interesting pattern while examining frequent patterns that included \texttt{stage2}. 
He identified the pattern \texttt{stage2-stage4}, where both stages represented relatively moderate states—neither extremely positive nor extremely negative (\autoref{fig: case1}-a2). \Ea was interested in how the transition between these stages might lead to subsequent behaviors.

\begin{figure}[h]
\centering
\includegraphics[width=\linewidth]{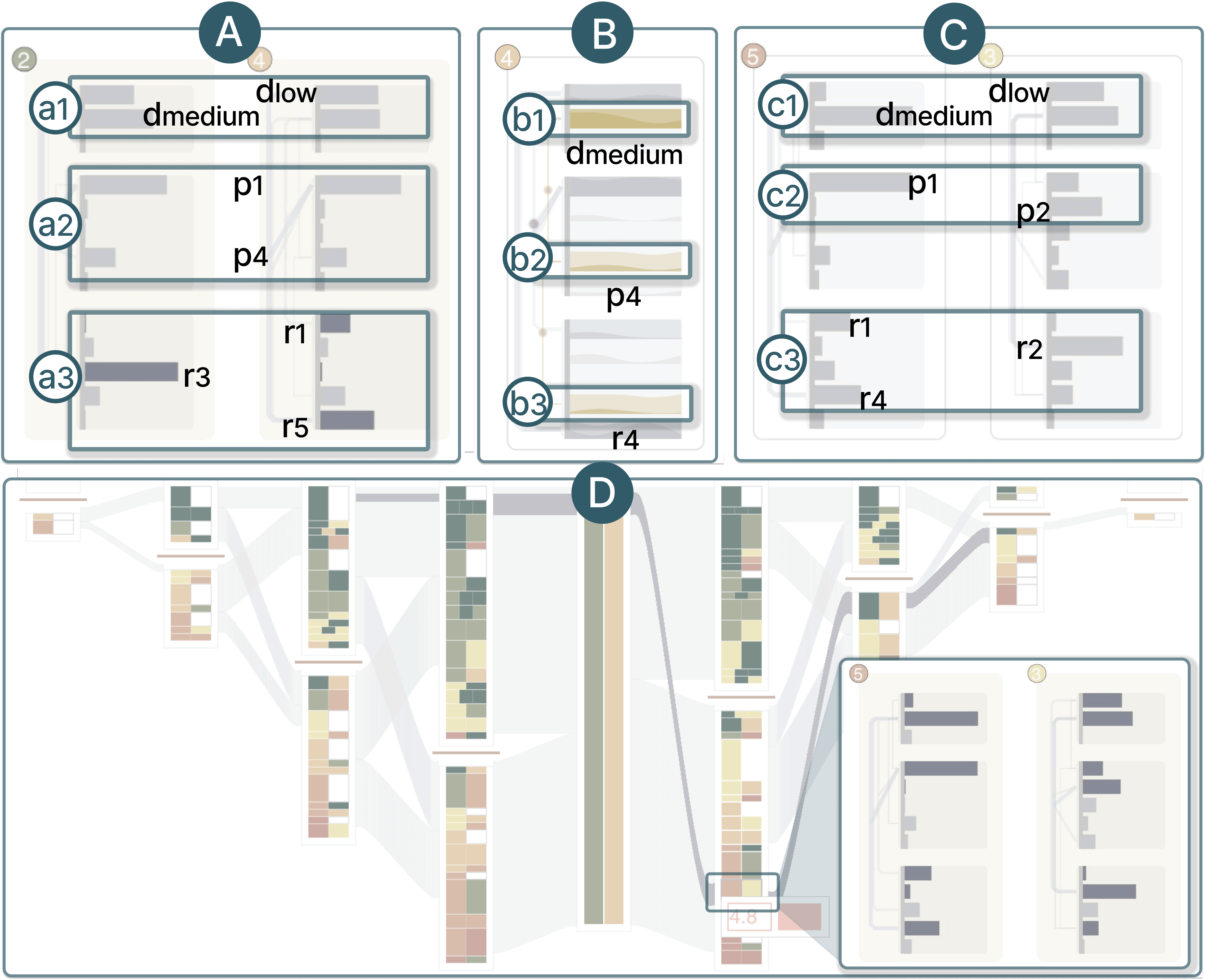}
\caption{Case II. (A-C) Understand the meanings of the behavior stage patterns \texttt{stage2-stage4} and \texttt{stage5-stage3}. (c1-c3) For example, from \texttt{stage5} to \texttt{stage3}, the depression decreased obviously, and proactive behaviors shifted from \texttt{p1} to \texttt{p2}, indicating gradual inactive seeking of social support, and reactive behaviors shifted from \texttt{r1} and \texttt{r4} to \texttt{r2}, suggesting a decrease in received social support. (B) Temporal mode of \texttt{stage4}. (D) Check the concrete stage sequences and hover over the stage pattern \texttt{stage5-stage3}.}
\vspace{-1em}
\label{fig: case2}
\end{figure}

\textbf{What is the implication of this pattern?} Specifically, his analysis of the two stages indicated a reduction in users' depression degree from \texttt{stage2} to \texttt{stage4}, as indicated by an increased ratio of low depression (\autoref{fig: case2}-a1) (\textbf{T1, T2}).
Furthermore, both stages showed active proactive behaviors dominated by \texttt{p1}, indicating users actively sought social support  (a2).
However, there was a difference in the number of received responses between the two stages, with more responses in \texttt{stage2} (a3).
To gain further insight, he switched to the temporal mode for exploring temporal information.
He found that the ratio of \texttt{p4} (representing that users were actively seeking social support) decreased (\autoref{fig: case2}-b2). 
Therefore, he knew that the frequency of seeking social support was actually decreasing in \texttt{stage4}, which naturally led to fewer responses.
The pattern \texttt{stage2-stage4} represented users' decreasing depression, accompanied by a reduced desire for social support and fewer obtained responses.
This raised the question of why subsequent negative behaviors increased despite users receiving sufficient support and successfully overcoming \mbox{their depression}.

\textbf{Why do the subsequent negative behaviors increase?} To understand the phenomenon, \Ea explored the subsequent behaviors of \texttt{stage2-stage4}.
He observed that the flow after \texttt{stage5-stage3} always linked the negative behaviors (\textbf{T5}) (\autoref{fig: case2}-D).
Therefore, \Ea thought that the subsequent behaviors would be negative after users experienced \texttt{stage2}, \texttt{stage4}, \texttt{stage5}, and \texttt{stage3} in order.
To understand \texttt{stage5-stage3}, he compared the stage visualizations (\textbf{T2}).
Specifically, he observed a reduction in depression level in \texttt{stage3} compared to \texttt{stage5} (\autoref{fig: case2}-c1).
Besides, he found that the ratio of \texttt{p1} obviously decreased while the ratio of \texttt{p2} increased (c2) from \texttt{stage5} to \texttt{stage3}, indicating a gradual shift towards an inactive posting situation for seeking social support.
Additionally, the amount of social support that users received was also decreasing (c3) (\textbf{T1}).
Next, \Ea connected \texttt{stage2-stage4} and \texttt{stage5-stage3} in the \progressionstageview.
He identified a remarkable increase in depression level in \texttt{stage5} compared to \texttt{stage4} (\textbf{T2}).

\textbf{Investigating behavior fluctuations.}
He summarized the observed patterns.
\textit{``Users' depression levels always fluctuated. During times of heightened depression, users actively sought social support and received more support. However, as their depression decreased, they became less proactive in seeking support, leading to a decrease in the received support.''} This case emphasizes that online communities can complement efforts to tackle depression, but a comprehensive approach involving timely offline interventions is necessary. 

\section{Expert Interview}\label{expert_interview}
\sidecomment{R3C1}
To assess the system's usability and effectiveness, we conducted one-on-one interviews with four new experts (\Pa-\Pd) who have not participated in our system design process before. 
\Pa is a professor with 10 years of experience, focusing on the research of emotional behaviors such as depression and anxiety, based in China. \Pb is a Ph.D. student \rui{majoring in social science} with 3 years of research experience, exploring how to leverage AI to analyze online posts for mental health studies, also studying in China. \Pc is another Ph.D. student \rui{majoring in social science} in China with 4 years of research experience, focusing on social computing and mental health. Finally, \Pd is a Ph.D. graduate \rui{from the Department of Communication} with 5 years of experience, specializing in computational social science and health communication, currently conducting research in the United States.
Our evaluation was conducted via Zoom with screen sharing. We began with a 30-minute introduction, covering the research background and system overview. Participants then had 15 minutes to familiarize themselves with the system independently. They used the settings in Case 1 to complete five tasks (taking 20 minutes). Afterward, they spent 15 minutes organizing their insights. Finally, a 10-minute semi-structured interview was held to gather feedback on the system's workflow, design, interactions, and potential improvements. The task list and interview questionnaire are in Appendix-F.

\indent\textbf{System Workflow.} 
All participants appreciated the usefulness of our system and its alignment with their traditional understanding of behaviors.
\Pa and \Pb emphasized the limitations of analyzing a few behavior variables in their previous study and praised \systemname for filling this gap.
Particularly, \Pc found the ability to substitute their own data in the system to be particularly appealing, enhancing their impression of the system workflow.
In addition, we also found that our system facilitates the quantitative study of complex user behaviors. 
Previously, experts built hypotheses from literature and social science theories and conducted quantitative research to verify them. The procedure was time-consuming and difficult to find new insights. 
With this tool, they can uncover numerous insightful patterns directly, expanding opportunities for subsequent quantitative research.

\indent\textbf{Visual Designs and Interactions.} All the experts found our visual design and interactions highly useful. They were able to comprehend the system's components after exploring it freely.
\Pd appreciated the reminder annotations at the top of each view. \Pa said that the automatic highlighting in the Behavior Cluster View was particularly helpful in locating corresponding behavior types. 
Besides, \Pa and \Pd expressed that behavioral differences could be identified intuitively through the \newSankey, allowing for an immediate understanding of the overall situation regarding positive and negative behaviors.
In addition, \Pa and \Pc liked the seamless exploration of stage patterns in the \progressionview, where hovering over a pattern displayed the corresponding stage visualizations in a tooltip. This feature eliminated the need to switch to the \progressionstageview frequently.

\indent\textbf{Suggestions.} 
\Pb and \Pc hoped to use this tool for interventions, suggesting the incorporation of individual-level information, such as user profiles and original content. \Pc and \Pd proposed we could add a note-taking feature to the system to assist them in exploration.

\section{Discussion}

\subsection{Design Implications.}
We summarized two main design implications based on the development process and evaluation feedback of our system.
\subsubsection{Link original data to high-level abstractions}
Maintaining a close connection with the original data is crucial when continuously abstracting data to higher levels, especially encompassing multiple layers of abstraction in our context (\ie, behavior variables, behavior types, behavior stages, and stage patterns).
For example, \Pb mentioned that \textit{``Although I can remember the general meaning of each stage, I still need to refer back to the original data (\ie, behavior variables) to interpret behavior stages or other high-level abstractions to ensure the accuracy of my conclusions.''}
Our system provides experts with a quick way to trace back to lower levels of data abstraction, all the way down to the original data. 
This is achieved through a series of interactions and visual linkages, such as highlighting and pop-up panels.

\subsubsection{Align visualization order across views}
Maintaining a consistent arrangement order of visualizations that convey the same information across different views is crucial. In our initial design of the stage visualization (\autoref{fig: stage}-B), experts found it challenging to align the order of the behavior types in a stage glyph in the \progressionstageview with that in the \behaviourclusterview.
\Pa stated, \textit{``This inconsistent order increased my information burden.''}
This highlights the importance for designers to consider not only local factors (\eg, the specific information conveyed by a visualization) but also to take a holistic perspective of the overall design (\eg, the structure and design in other views).

\subsubsection{Applicability and Generalizability}
In addition to supporting the analysis of depression behaviors, our system can also be adapted to explore other mental health assessments, such as anxiety or schizophrenia~\cite{chung22mental} in terms of temporal evolution. 
For example, \Pa mentioned, \textit{``I can utilize anxiety levels as an indicator of a user's mental health status. Anxiety disorders represent the most prevalent category of mental disorders, and several models currently exist to predict anxiety based on users' posts.''}
Furthermore, our system can also inspire various studies on behavior progression, such as helping define user roles in online communities to optimize user experiences and foster long-term community growth.
\Pc mentioned that our system could help them explore intricate temporal sequences and define user roles more subtly and comprehensively, compared with the prior studies~\cite{welser2011finding, yang2019seekers}. 
Finally, our pipeline can also be applied to other domains involving the analysis of progression in multidimensional variables, such as career development~\cite{wang2022seek}. 
For instance, our pattern-centric Sankey diagram in the \progressionview can help experts uncover important stages in the career trajectories of a group of users.

\subsection{Limitations and Future Work.}
Despite the contributions of this work, we acknowledge several limitations and highlight future opportunities.

\subsubsection{Algorithm performance and robustness} The current model~\cite{poswiata-perelkiewicz-2022-opi} used to predict depression and classify posts into three levels (\ie, not depressed, moderately depressed, and severely depressed) may present uncertainties, achieving an accuracy of 0.7 on the dev dataset and 0.66 on the test dataset as reported in the previous paper.
However, it is important to emphasize that this tool is intended for analyzing the behavior patterns of online users rather than diagnosing individual mental health conditions or developing precise treatment plans. As such, experts acknowledge the model's accuracy and still regard it as a valuable tool for uncovering initial scientific insights. 
Furthermore, our main contribution is not in developing algorithms for extracting variables like depression or toxicity, but in offering a pipeline for analyzing behavior variables.
This pipeline provides specific procedures to assist experts in conducting online behavior data analysis.

\subsubsection{Determination of cluster numbers} The cluster number setting in the Behavior Cluster View might introduce uncertainty in constructing behavior events.
Our system provides visual aids to help users select a cluster number and reduce uncertainty. However, there is still a challenge in fully exploring the data patterns and trends due to the complexity and volume of data. Future improvements include highlighting newly emerged clusters and enabling experts to merge similar clusters, further reducing uncertainty.

\subsubsection{Triggers for understanding in behavior transition} Two experts (\Pb and \Pc) during the interview presented a strong interest in understanding contextual information, such as triggers (\ie, specific cues that prompt a change in behavior) for transitions between behavior stage patterns. In the future, we can extract triggers~\cite{remes2021bio} from social media data and display them in the \newSankey.

\subsubsection{Specificity of findings and caution in broader application} Although using extracted stage patterns for interventions offers new possibilities for research, it is important to note that our findings are specific to the social media environment and the studied population. 
Generalizing these results to other contexts or populations should be addressed with caution, considering potential differences and nuances. 

\section{Conclusion}
In this work, we propose \systemname, a visual analytics system to help social scientists analyze user behavior progression in OMHCs. The system uses an automatic mining pipeline to extract behavior stages and identify crucial stage patterns for long-term mental health. Interactive visualizations enable comparisons between behavior stages, and the aligned Sankey diagram shows detailed behavior evolution before and after specific stage patterns. Two case studies and the expert interview have demonstrated the system's efficacy and usability.
\section*{Acknowledgments}
We sincerely thank the reviewers for their insightful feedback. This work was also supported in part by the U.S. National Science Foundation under Grant No. IIS-2427770.
\bibliographystyle{IEEEtran}
\bibliography{ref}
\begin{IEEEbiography}[{\includegraphics[width=1in,height=1.25in,clip,keepaspectratio]{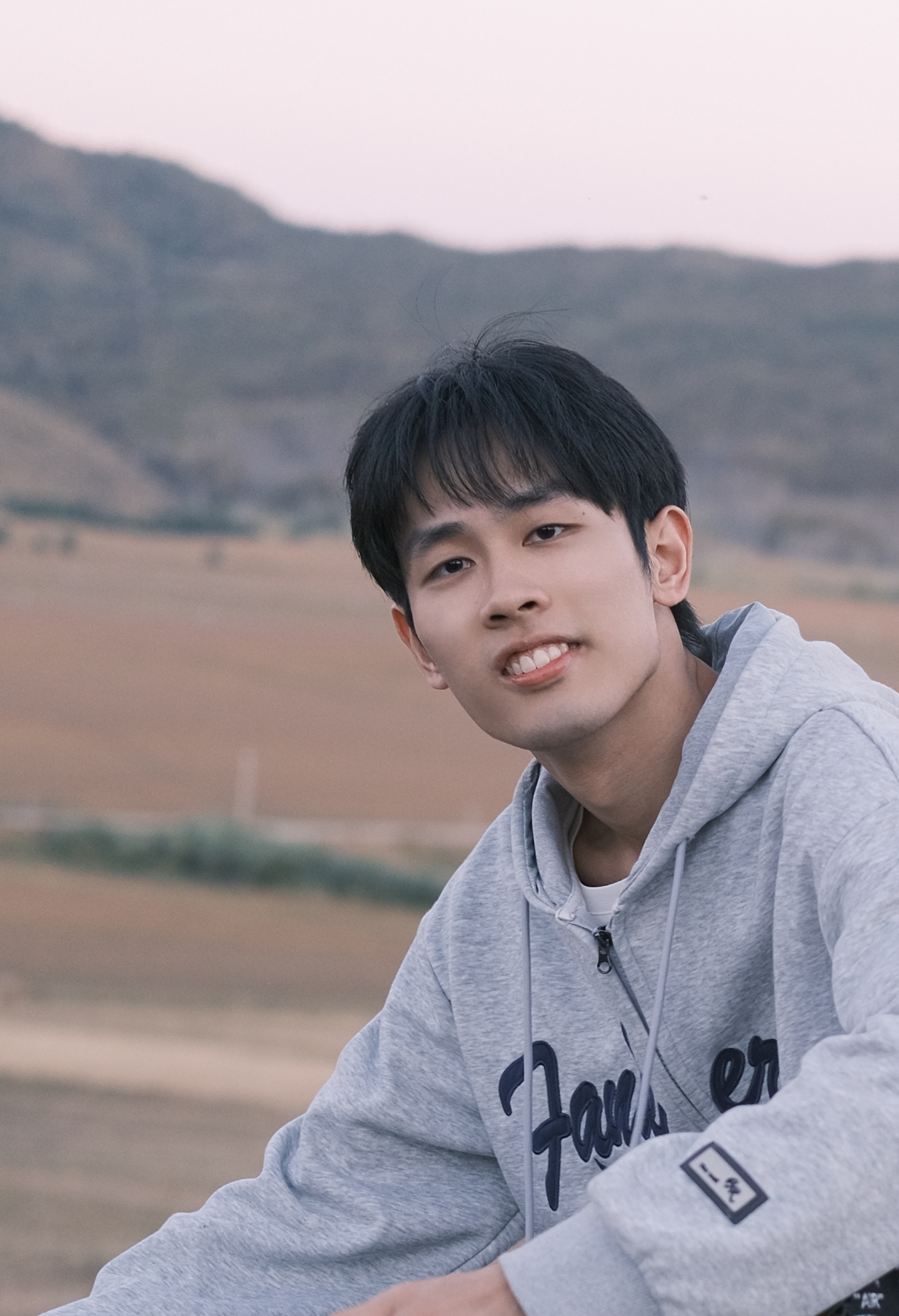}}]{Rui Sheng} is a Ph.D. candidate at HKUST (Hong Kong University of Science and Technology) VisLab under the supervision of Professor Huamin Qu. His research focuses on data visualization, human-AI collaboration, and decision-making, particularly in critical domains such as healthcare. He is dedicated to developing effective approaches for making informed decisions based on sequential data. More information: \url{https://dylansheng.github.io/}.
\end{IEEEbiography}
\vspace{-40pt}
\begin{IEEEbiography}[{\includegraphics[width=1in,height=1.25in,clip,keepaspectratio]{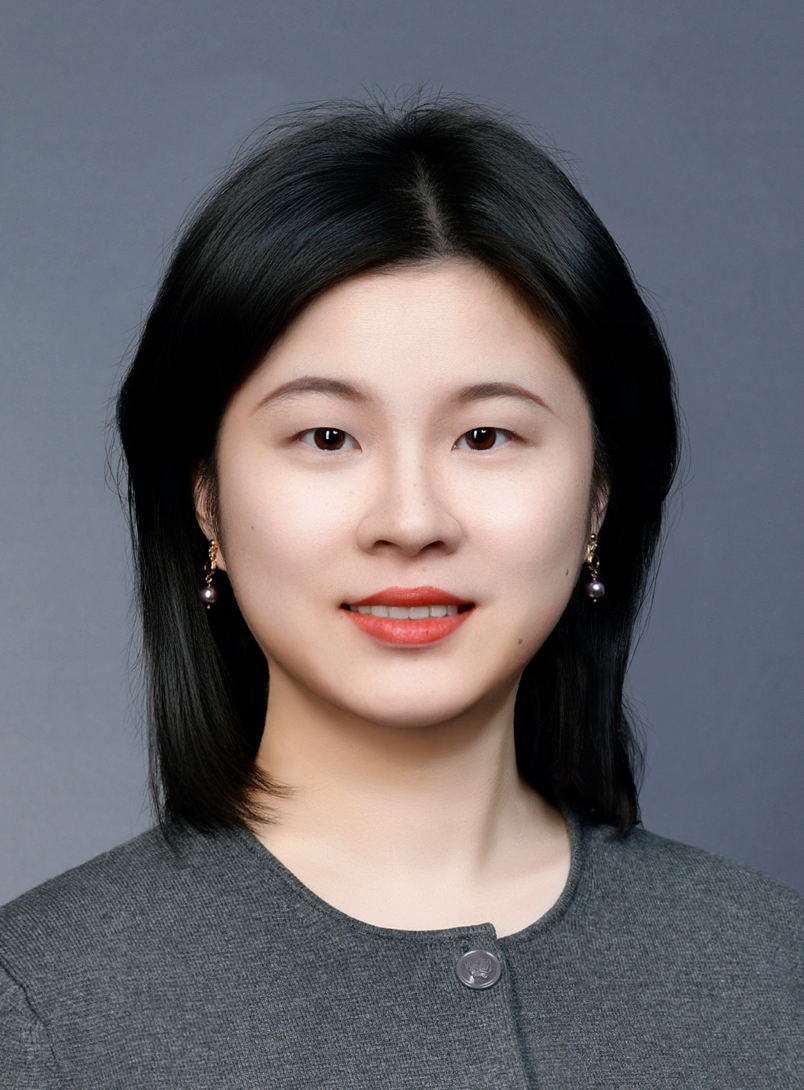}}]{Yifang Wang} is a tenure-track Assistant Professor at the Department of Computer Science at Florida State University. 
Her research lies at the intersection of Data Visualization, Human-AI Collaboration, and Human-Computer Interaction. She develops novel VIS and HAI techniques and tools that support not only general data-driven tasks but also domain-specific applications in areas such as the science of science and innovation, science policy-making, public health, and more. 
More information: \url{https://wangyifang.top}.
\end{IEEEbiography}
\vspace{-40pt}
\begin{IEEEbiography}[{\includegraphics[width=1in,height=1.25in,clip,keepaspectratio]{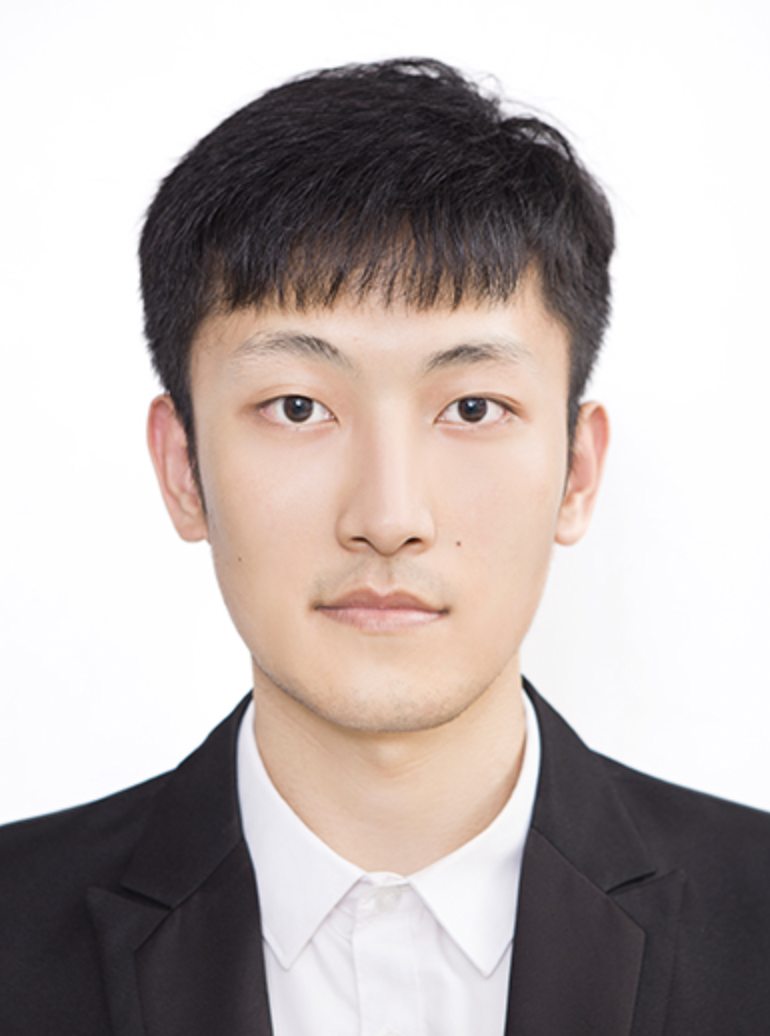}}]{Xingbo Wang} is currently a postdoctoral associate at the Weill Cornell Medical School of Cornell University. He received his Ph.D. degree in 2022. He obtained a B.E. degree from Wuhan University, China in 2018. His research interests include human-computer interaction (HCI), data visualization, natural language processing (NLP), and multi-modal analysis. More information, \url{https://andy-xingbowang.com/}.
\end{IEEEbiography}
\vspace{-40pt}
\begin{IEEEbiography}[{\includegraphics[width=1in,height=1.25in,clip,keepaspectratio]{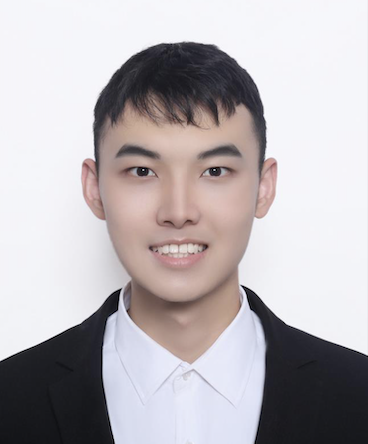}}]{Shun Dai} is a currently master's student at The University of Tokyo, passionately pursuing his research in the captivating field of virtual reality. Alongside his academic endeavors, he actively engages as a self-funded independent researcher, delving into innovative frontiers of human-computer interaction in his spare time.
\end{IEEEbiography}
\vspace{-40pt}
\begin{IEEEbiography}[{\includegraphics[width=1in,height=1.25in,clip,keepaspectratio]{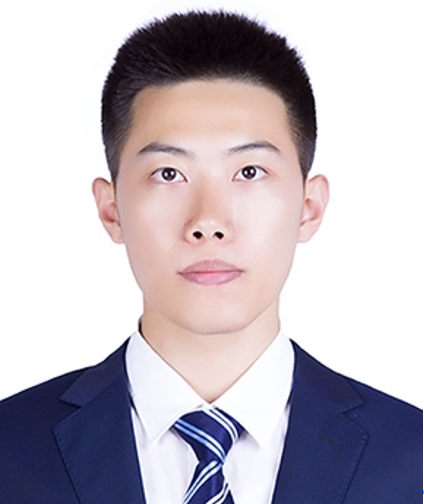}}]{Qingyu Guo} is currently a Ph.D. candidate from Hong Kong University of Science and Technology. His advisor is Prof. Xiaojuan Ma. Specifically, he is interested in social computing and human-centered machine learning. He immerses himself in understanding human communication and enhancing their capabilities in social context, especially for online social support and collaboration. More information: \url{https://qingyuguo.github.io/}.
\end{IEEEbiography}
\vspace{-40pt}
\begin{IEEEbiography}[{\includegraphics[width=1in,height=1.25in,clip,keepaspectratio]{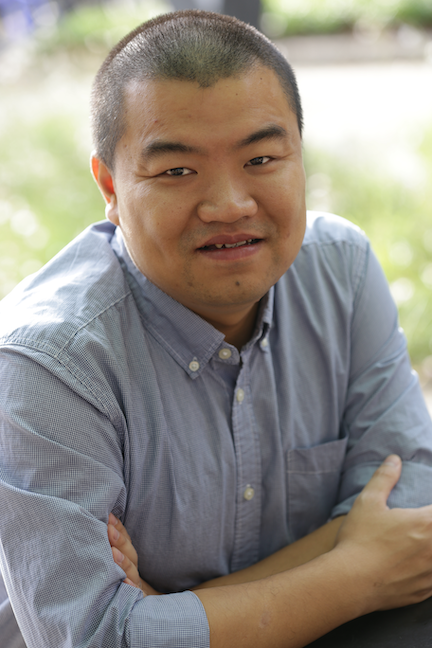}}]{Tai-Quan Peng} is currently a professor in the Department of Communication, Michigan State University.
His research interest includes computational social science, health communication, mobile analytics, and political communication. He is particularly interested in unraveling the structure \& dynamics of human communication phenomena with various computational methods (e.g., network modeling, text mining, temporal and sequential modeling). More information: \url{https://winsonpeng.github.io/}.
\end{IEEEbiography}
\vspace{-40pt}
\begin{IEEEbiography}[{\includegraphics[width=1in,height=1.25in,clip,keepaspectratio]{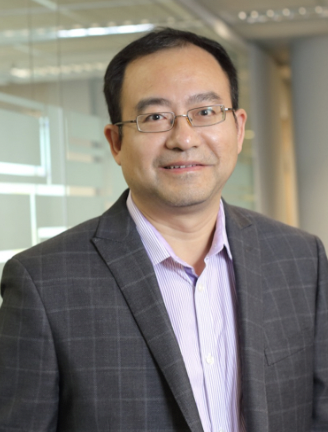}}]{Huamin Qu} is the dean of the Academy of Interdisciplinary Studies, head of the Division of Emerging Interdisciplinary Areas, and a chair professor in the Department of Computer Science and Engineering at HKUST. He obtained a BS in Mathematics from Xi’an Jiaotong University, China, an MS, and a PhD in Computer Science from the Stony Brook University. His main research interests are in visualization and human-computer interaction, with focuses on urban informatics, social network anal- ysis, Elearning, text visualization, and explainable artificial intelligence. More information: \url{http://huamin.org/}.
\end{IEEEbiography}
\vspace{-40pt}
\begin{IEEEbiography}[{\includegraphics[width=1in,height=1.25in,clip,keepaspectratio]{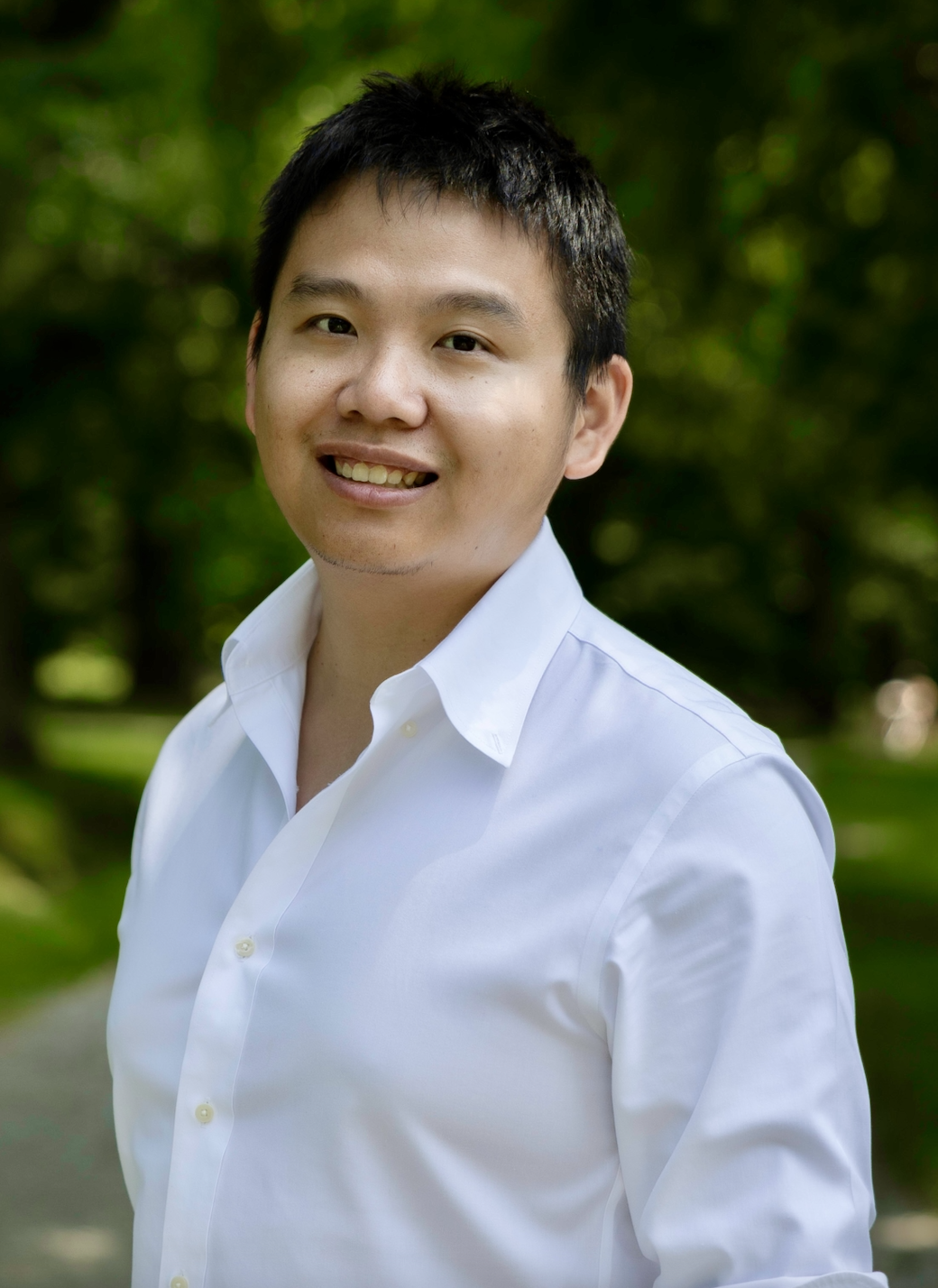}}]{Dongyu Liu}  is an assistant professor in the Department of Computer Science at the University of California, Davis, where he directs the Visualization and Intelligence Augmentation (VIA) Lab. His research develops visualization-empowered human-AI teaming systems for decision-making, focusing on sustainability and healthcare. He was a Postdoctoral Associate at MIT and earned his Ph.D. from HKUST. More information: \url{http://dongyu.tech/}.
\end{IEEEbiography}

\end{document}